\documentclass[12pt]{article}
\usepackage{amsmath,amssymb,amsfonts,amscd,amsxtra}
\usepackage{latexsym}
\usepackage{bm}

\input{epsf.sty}
\usepackage{epsfig}
\usepackage{graphicx,color}
\usepackage{graphics}
\usepackage{subfigure}
\usepackage{appendix}
\usepackage{wrapfig, blindtext}
\usepackage[top=1in, bottom=1in, left=1in, right=1in]{geometry}

\renewcommand{\theequation}{\thesection.\arabic{equation}}

\newtheorem{thm}{Theorem}[section]

\newtheorem{lem}[thm]{Lemma}
\newtheorem{prop}[thm]{Proposition}
\newtheorem{defn}{Definition}
\newtheorem{rem}{Remark}
\newtheorem{asump}{Assumption}

\renewcommand{\Im}{{\mbox{Im}}}
\renewcommand{\Re}{{\mbox{Re}}}

\newcommand{\norm}[1]{\left\Vert#1\right\Vert}
\newcommand{\abs}[1]{\left\vert#1\right\vert}

\newcommand{\be}{\begin{equation}}
\newcommand{\ee}{\end{equation}}

\newcommand{\R}{\mathbf{R}}
\newcommand{\Ao}{\mathcal{A}}
\newcommand{\B}{\mathcal{B}}

\newcommand{\Lo}{\mathcal{L}}

\newcommand{\qed}{\hfill \ensuremath{\square}}

\renewcommand\appendix{\par
  \setcounter{section}{0}
  \setcounter{subsection}{0}
  \setcounter{figure}{0}
  \setcounter{table}{0}
  \renewcommand\thesection{Appendix \Alph{section}}
  \renewcommand\theequation{\Alph{section}.\arabic{equation}}
  \renewcommand\thefigure{\Alph{section}.\arabic{figure}}
  \renewcommand\thetable{\Alph{section}.\arabic{table}}
  \renewcommand\thethm{\Alph{section}.\arabic{thm}}
}

\numberwithin{equation}{section}

\date{}

\title{Mathematical theory for topological photonic materials in one dimension}

\author{
Junshan Lin \thanks{\footnotesize Department of Mathematics and Statistics, Auburn University, Auburn, AL 36849 (jzl0097@
auburn.edu). Junshan Lin was partially supported by the NSF grant DMS-2011148.}
 \; and Hai Zhang\thanks{\footnotesize 
Department of Mathematics, 
 HKUST,  Clear Water Bay, Kowloon, Hong Kong SAR, China (haizhang@ust.hk). Hai Zhang is partially supported by Hong Kong RGC grant GRF 16304517 and GRF 16306318.}}

\begin{document}
\maketitle

\begin{abstract}
This work presents a rigorous theory for topological photonic materials in one dimension. The main focus is on the existence and stability of interface modes that are induced by topological properties of the bulk structure. For a general 1D photonic structure with time-reversal symmetry, the associated Zak phase (or Berry phase) may not be quantized. We investigate the existence of an interface mode which is induced by a Dirac point upon perturbation.   
Specifically, 
we establish conditions on the perturbation which guarantee the opening of a band gap around the Dirac point and the existence of an interface mode. 
For a periodic photonic structure with both time-reversal and inversion symmetry, the Zak phase is quantized, taking only two values $0, \pi$. We show that the Zak phase is determined by the parity (even or odd) of the Bloch modes at the band edges. 
For a photonic structure consisting of two semi-infinite systems on the two sides of an interface with distinct topological indices, we show the existence of an interface mode inside the common gap. The stability of the mode under perturbations is also investigated. 
Finally, we study resonances for finite topological structures.
Our results are based on the transfer matrix method and the oscillation theory for Sturm-Liouville operators. The methods and results can be extended to general topological Sturm-Liouville systems in one dimension. 
\end{abstract}


\setcounter{equation}{0}
\setlength{\arraycolsep}{0.2em}

\section{Introduction}

Topological insulator is a phase of matter that 
conducts electrons on its edge or interface without backscattering.
The underlying protected edge/interface mode is robust at the presence of large impurities,
which prevent the degradation of device performance due to fabrication imperfections.
Tremendous progress has been made in the past several decade 
in the studies of topological insulators and quantum topological materials in general in electron systems \cite{Hasan-Kane-10, bernevig-13}. 
In recent years, there have been increasing interests in exploring the analogue of the quantum topological materials for periodic photonic/phononic band gap materials \cite{HR-08, Khanikaev-12, Lu-14, Ozawa-19, raghu-08, Rechtsman-13}. 

From the mathematical point of view, there are several important issues in the studies of topological materials. The first one is concerned with the Dirac points of the band structure for the toplogical material.
Dirac points are special vertices located at the Brillouin zone corners when two bands in the spectrum touch in a linear conical fashion 
and degeneracies occur for the corresponding Bloch modes \cite{HR-08}.
We refer to \cite{Fefferman-Lee-Thorp-Weinsein-17} for the rigorous mathematical studies of Dirac points in 1D periodic Schr\"{o}dinger operator with double-well potential and
\cite{Fefferman-Weinsein-12} for the construction of Dirac points for Schr\"{o}dinger operator with the Honeycomb lattice potentials in 2D.
In general, a topological phase transition takes place near the Dirac point and interesting physics phenomena occurs as a result. This is exemplified in  
photonic graphene and subwavelength resonantors in \cite{Ablowitz-Zhu-12, ammari-20-2, Fefferman-Weinsein-14, Lee-Thorp-Weinstein-Zhu-19} and 
 references therein.
The second one is the existence of interface modes (also called edge modes or edge states) that are supported at the interface of two structures with distinct topological invariants. This is typically formulated as the so-called bulk-edge correspondence, which formally states that the bulk index is equal to the edge index. The former is a topological quantity that can be computed from the bulk media, while the latter is related to the number of edge modes supported by the structure.
A variety of tools have been developed for the study of the bulk-edge correspondence in different settings, including K-theory, functional
analysis, and microlocal analysis, etc \cite{bal-17, bal-19, druout-20, druout-20-3, EGS-05, Graf-02, hatsugai-93, Kellendonk-04, Shapiro, Taarabt-14}. The third one is the stability of the interface modes supported by the topological materials. Such modes are
``topologically protected" in the sense that they are stable against the system perturbations that are not necessarily small; See, for instance, \cite{Fefferman-Lee-Thorp-Weinsein-17} and \cite{ammari-20-3} for the mathematical investigation of stability for edge mode in 1D Schr\"{o}dinger system and subwavelength resonators respectively.

In this paper, we study one-dimensional photonic structures with time-reversal symmetry.
The corresponding periodic differential operator is defined by
\begin{equation} \label{eq-photonic}
    \Lo\psi = -\frac{1}{\varepsilon(x)} \dfrac{d}{dx} \left( \frac{1}{\mu(x)}\frac{d\psi}{dx}\right) \quad\mbox{for} \; x\in\mathbf{R},
    \end{equation}
where the permittivity $\varepsilon(x)$ and the permeability $\mu(x)$ are two positively valued piecewisely continuous functions with period 1:
\begin{equation}
    \varepsilon(x)= \varepsilon(x+1), \quad    \mu(x)= \mu(x+1).
\end{equation}
We aim to provide a rigorous mathematical theory for the given one-dimensional topological structure, especially on the existence and stability of the interface modes.
Based on the transfer matrix method, we characterize the Dirac points of the structure precisely.
We present explicit conditions for the perturbation of the structure
so that a band gap can be opened near the Dirac point and an interface mode can be generated.
For structures with additional inversion symmetry, we provide explicit formulas for the Berry phase, which is also called the Zak phase for one-dimensional structures. 
In this scenario, the Zak phase is closely related to the parity of the Bloch modes at the band edges. It is quantized by taking the value of $0$ or $\pi$ only and hence becomes a natural bulk topological index. 
We establish the existence and investigate the stability of interface modes when two semi-infinite periodic structures attain distinct topological indices.
Furthermore, we study the resonances for finite topological structures, for which the eigenvalues are complex-valued and the eigenmodes increase exponentially at infinity.

We mention several closely related work \cite{Fefferman-Lee-Thorp-Weinsein-17, druout-20-1, druout-20-2}, where one-dimensional Schr\"{o}dinger equations with periodic potentials are studied. It is shown in \cite{ Fefferman-Lee-Thorp-Weinsein-17} that for a class of background periodic Schr\"{o}dinger operators with Dirac points, localized edge states can be induced via small and adiabatic modulation of the periodic
potentials with a domain wall, and the bifurcation of these states are associated with the discrete eigenmodes of an effective Dirac operator. The studies are based on the multiple-scale analysis and Lyapunov-Schmidt reduction technique. In \cite{druout-20-1}, the author studies a topological system where the background periodic Schr\"{o}dinger operator is perturbed by a small and 
adiabatic dislocation. It is shown that all the edge states of the dislocated
system are associated with the eigenmodes of an effective Dirac operator. Moreover, full asymptotic expansions of the eigenpairs are derived. 
In \cite{druout-20-2}, the bulk-edge correspondence is rigorously established for a family of operators, wherein each operator corresponds to a dislocation of the background periodic Schr\"{o}dinger operator. It is proved that certain edge index is equal to the bulk index given by the Chern number of the Bloch
eigenbundle below the band gap. This result justifies the heuristic statement that the edge modes are ``topologically protected" at the level of a family of operators. 
In addition, we refer to
\cite{ammari-20-1, ammari-20-3} for the studies on topologically protected edge states in a one-dimensional chain of subwavelength resonators in three dimensions. 

The rest of the paper is organized as follows. In Section \ref{sec:periodic_structure}, 
the band structure theory for the 1D periodic differential operators is recalled.  Furthermore, Dirac points at the Brillouin zone corners are investigated and the Zak phase over the band structure is computed.
Section \ref{sec:perturb_sys} studies the perturbation of a general time-reversal symmetric structure with a Dirac point and the existence of an interface mode for the perturbed system.
Section \ref{sec:inv_sym} focuses on time-reversal symmetric structures that attain inversion symmetry.
The existence of an interface mode that is predicted by the bulk topological indices and its stability under perturbations that are not necessarily small are established. Finally, the studies of resonances for finite topological structures is provided 
in Section \ref{sec-resonance}.

\section{Band structure, Dirac point and Zak phase for the periodic structure}\label{sec:periodic_structure}
In this section, we recall the band structure theory for the spectrum of the periodic differential operator $\Lo$. Furthermore, we investigate Dirac points at the corners of the reduced Brillouin zone and 
compute the Zak phase.

\subsection{Spectrum of the operator $\Lo$}
The spectrum of the operator $\Lo$ can be characterized using the Floquet-Bloch theory and the transfer matrix.
For completeness we collect several key results in this section.
The readers are referred to \cite{Kuchment-12} for more details about the Floquet-Bloch theory for periodic differential operators.

Throughout, we denote $L^2(\R)$ for the Hilbert space equipped with the inner product
$$
(\psi, \phi) = \int_{\R} \varepsilon(x) \psi(x) \bar{\phi}(x) dx, 
$$
and we denote $X$ for the Hilbert space $L^2[0, 1]$ equipped with the inner product
$$
(u, v)=\int_{0}^1 \varepsilon(x) u(x) \bar{v}(x) dx. 
$$
Let $\B = [-\pi, \pi]$ be the Brillouin zone. The reduced Brillouin zone is $[0, \pi]$. For each Bloch wavenumber $k \in \B$, we consider the following 
one-parameter family of Floquet-Bloch
eigenvalue problem  
\begin{equation} \label{eigen-k}
     \Lo \psi(x)= E \psi(x) \quad x\in\R, \quad \psi(x+1)= e^{ik}\psi(x)
\end{equation}
in the function space
$$
L^2_k = \{u \in L^2_{loc}: u(x+1)= e^{ik} u(x) \}. 
$$
The eigenvalue problem (\ref{eigen-k}) is self-adjoint and has a discrete set of real eigenvalues
$$
E_1(k) \leq E_2(k) \leq \cdots \leq E_j(k) \leq \cdots. 
$$
The eigenfunction associated with the eigenvalue $E_j(k)$ is called the $j$-th Bloch mode. 

\begin{lem} \label{lem-11}
\begin{enumerate}
\item [(1)]
The function $E_j(k)$, also called the dispersion relation of the $j$-th band, is Lipschitz continuous with respect to $k\in \B$. 

\item [(2)]
$E_j(k) = E_j(-k)$ holds for each $k\in \B$. Moreover, $E_j(k)$ can be extended to a periodic function in $k$ with period $2\pi$, i.e.  $E_j(k)= E_j(k+ 2\pi)$;   

\item [(3)]
$E_1(k) \geq 0$ for each $k\in \B$. In addition, $E_1(0) =0$, and the corresponding Bloch mode is a constant function. 
\end{enumerate}
\end{lem}
\noindent\textbf{Proof.} The first two statements (1) and (2) are standard results in the spectral theory of periodic differential operators. The readers may refer to chapter XIII, section 16 in \cite{Reedsimon} for the case of Schr\"{o}dinger operators with periodic potentials. The extension to the operator $\mathcal{L}$ considered in this paper is straightforward. The last statement (3) is trivial. \qed

\medskip

For each integer $j$, let 
$$
E_j^-=\min \{ E_j(k): k\in \B \}, \quad  E_j^+=\max \{ E_j(k): k\in \B\}. 
$$
Then the entire spectrum spectrum of the operator $\Lo$ on $L^2(\R)$ is given by
$$
\sigma(\Lo) = \bigcup\limits_{j\geq 1} \, [E_j^-, E_j^+],
$$
which corresponds to the essential part of the spectrum. The spectrum forms a band gap if $E_j^+ < E_{j+1}^-$ for some $j$.

The band structure of the spectrum can be characterized using the transfer matrix method. 
To this end, for each $E\in \mathbf{R}$ we let 
$\psi_{E,1}$ and $\psi_{E, 2}$ to be the unique solution to the following problems:
\begin{align}
(\Lo - E)   \psi_{E,1} =0, \quad \psi_{E,1}(0)= 1, \,\, \frac{1}{\mu(0)}\psi_{E,1}'(0)=0, \\
(\Lo-E)   \psi_{E,2} =0, \quad \psi_{E,2}(0)= 0, \,\, \frac{1}{\mu(0)}\psi_{E,2}'(0)=1.  
\end{align}
Define
\be  \label{eq-psi}
\Psi_E (x) = (\Psi_{E, 1}(x), \Psi_{E, 2}(x) )= 
\begin{pmatrix} 
\psi_{E,1}(x) & \psi_{E,2}(x)   \\
\frac{1}{\mu(x)}\psi_{E,1}'(x)  & \frac{1}{\mu(x)}\psi_{E,2}'(x)
\end{pmatrix},
\ee
It is clear that $\Psi_E (x)$ solves the the initial value problem
\be
\frac{d}{dx}\Psi_E (x) = J(B + EW)\Psi_E(x), \quad  \Psi_E(0)= Id, 
\ee
where 
\be
J= 
 \begin{pmatrix}
 0 & 1 \\
 -1 & 0\\
 \end{pmatrix}, \quad 
 B= B(x)=\begin{pmatrix}
 0 & 0 \\
0 & \mu(x)\\
 \end{pmatrix}, \quad 
 W= W(x)=\begin{pmatrix}
\varepsilon(x) & 0 \\
0 & 0\\
 \end{pmatrix}.
\ee

\begin{rem}
In the above , $\frac{1}{\mu(0)}\psi_{E,1}'(0)$ means either $\lim_{x \to 0^+}\frac{1}{\mu(x)}\psi_{E,1}'(x)$ or $\lim_{x \to 0^-}\frac{1}{\mu(x)}\psi_{E,1}'(x)$. The two terms are equal even when $x=0$ is a point of discontinuity of $\varepsilon$ and $\mu$ by the continuity of the flux. Similar notations will be used throughout the paper.  
\end{rem}

Let $M(E)=\Psi_E (1) $, which is called the monodromy matrix. The eigenvalues of $\Lo$ in $L^2_k$  can be characterized by the eigenvalues of the matrix-valued function $M(E)$. 

\begin{lem} \label{lem-eigen1}
\begin{enumerate}
\item [(1)]
If $(\psi, E)$ is an eigenpair in $L^2_k$, then $(\psi(0), \frac{1}{\mu(0)} \psi'(0))^T$ is an eigenvector of the matrix $M(E)$ with the corresponding eigenvalue $e^{ik}$. 
\item [(2)]
If the matrix $M(E)$ has an eigenpair $(e^{ik}, (a_1, a_2)^T)$, then $\psi(x) = a_1 \psi_{E,1}(x) + a_2 \psi_{E,2}(x)$ is a Bloch mode of $\Lo$ in $L^2_k$. 
\end{enumerate}
\end{lem}

\noindent\textbf{Proof.}  Assume that $(\psi, E)$ is an eigenpair in $L^2_k$. We can write
\begin{equation}\label{psi_exp}
\psi(x) = a_1 \psi_{E,1}(x) + a_2 \psi_{E,2}(x), 
\end{equation}
where $a_1 = \psi(0), a_2 = \frac{1}{\mu(0)} \psi'(0)$. Then
\begin{align}
\psi(1) &= a_1 \psi_{E,1}(1) + a_2 \psi_{E,2}(1)  = e^{ik} \psi(0) = e^{ik} a_1,  \\
\frac{1}{\mu(1)} \psi'(1) &= \frac{1}{\mu(1)} \big(  a_1 \psi_{E,1}'(1) + a_2 \psi_{E,2}'(1) \big) = e^{ik} a_2,
 \end{align}
or equivalently, 
$$
M(E) a = e^{ik} a, \quad \mbox{where} \; a=(a_1, a_2)^T.
$$ 
It follows that the matrix $\Psi_E(1)$ has eigenvalue $e^{ik}$ and the associated eigenvector is $a$. 
On the other hand, assuming that $M(E)$ has an eigenpair $(e^{ik}, a)$.
We construct $\psi$ as in \eqref{psi_exp}. It is straightforward to show that $\psi$ is a Bloch mode of $\Lo$ in $L^2_k$.  \qed


\begin{lem}\label{lem-detE}
det $M(E)$ =1.
\end{lem}
\noindent\textbf{Proof.}  For each fixed $E$, consider $f(x)$ = det $\Psi_E(x)$.  A direct calculation shows that $f'(x)=0$. As a result, det $\Psi_E(x)$ is independent of $x$. Note that $\Psi_E(0)=Id$, we get the desired result immediately. \qed  \\

Define 
$$
D(E)= Tr M(E) = \psi_{E, 1}(1) + \frac{1}{\mu(1)}\psi_{E,2}'(1).
$$
$D(E)$ is called the discriminant of the $M(E)$. It is clear that $D$ is real valued. The two eigenvalues of the matrix $M(E)$ are given by
\begin{equation}\label{eq-lambda1_lambda2}
\lambda_{E, 1}= \frac{D(E) - \sqrt{D(E)^2 -4}}{2}, \quad \lambda_{E, 2} = \frac{D(E) + \sqrt{D(E)^2 -4}}{2}.
\end{equation}
If $|D(E)| \leq 2$, then $ \lambda_{E, 1}$ and $ \lambda_{E, 2}$ are conjugate pair with $|\lambda_{E, 1}|= |\lambda_{E, 2}|=1$. 
It follows from Lemma \ref {lem-eigen1} that $E\in \sigma(\Lo)$. On the other hand, 
if $|D(E)| > 2$, then both $ \lambda_{E, 1}$ and $ \lambda_{E, 2}$ are real numbers satisfying
$|\lambda_{E, 1}| <1<|\lambda_{E, 2}|$ or $|\lambda_{E, 2}| <1<|\lambda_{E, 1}|$. 
In this case, $E\not\in \sigma(\Lo)$ and it lies in the band gap. In summary, we have the following lemma for the spectrum of $\Lo$ and the discriminant $D(E)$.

\begin{lem}  \label{lem-12}
The real number $E\in \sigma(\Lo)$ if and only if $|D(E)| \leq 2$.
\end{lem}

Let
$$
\mathcal{S}= \{E \in \mathbf{R}: |D(E)| < 2\} \quad \mbox{and} \quad \mathcal{I}= \{E \in \mathbf{R}: |D(E)| > 2\}.
$$
Then the following lemma holds. We refer to Theorem 1.6.1 in \cite{brown-2013} for its proof.
\begin{lem} \label{lem-pre1} 
\begin{enumerate}
\item [(1)]
The function $D(E)$ is strictly monotonic on each subinterval of $\mathcal{S}$.
\item [(2)]
For $E \in \mathbf{R}$, $D(E)= 2$, $D'(E)=0$ holds if and only if
$M(E)= Id$. In this case, $D''(E) <0$. 
\item [(3)]
$D(E)= -2$, $D'(E)=0$ holds if and only if 
$ M(E)= -Id$.  In this case, $D''(E) >0$.
\end{enumerate}
\end{lem}

From the above lemmas, the band structure of the spectrum of the operator $\Lo$ can be characterized in the theorem below. 


\begin{thm} \label{thm-bandstructure}
\begin{enumerate}

\item [(1)]
The following inequalities hold for the spectrum $\sigma(\Lo) = \bigcup\limits_{j\geq 1} [E_j^-, E_j^+]$:
$$
0=E_1^- < E_1^+ \leq E_2^- < E_2^+ \leq  E_3^- < E_3^+ \cdots 
$$
\item  [(2)] 
The dispersion relation $E_j = E_j(k)$ can be obtained by solving the equation
\begin{equation}\label{eq-disp}
2 \cos k = D(E)
\end{equation}
for $k\in \B $ and $E \in [E_j^-,  E_j^+]$. 

\item  [(3)]
$E_j(k)$ are strictly monotonic on each of the half Brillouin zone $(-\pi, 0)$ and $(0, \pi)$.

\item  [(4)]
For each $j \geq 1$, we have either
$$
E_j^+ = \max \{ E_j(k): k\in \B\} = E_j(0), \,\,  E_{j+1}^- = \min \{ E_{j+1}(k): k\in \B\}= E_{j+1}(0),
$$
or
$$
E_j^+ = \max \{ E_j(k): k\in \B\} =E_j(\pi), \,\,  E_{j+1}^- = \min \{ E_{j+1}(k): k\in \B\}=E_{j+1}(\pi).
$$

\item  [(5)]
If $D(E^*)= \pm 2$ and $D'(E^*)=0$, then
$E^*= E_j^+ = E_{j+1}^-$ for some $j\geq 1$. Moreover, $E^*= E_j(0)=E_{j+1}(0)$ if $D(E^*)= 2$ and $E^*= E_j(\pi)=E_{j+1}(\pi)$ if $D(E^*)= -2$. 
\end{enumerate}
\end{thm}

\noindent\textbf{Proof.}
(1). By Lemmas \ref{lem-11} and  \ref{lem-12}, we see that $|D(E)| >2$ for $E<0$, thus $\mathcal{S} \subset (0, \infty)$. 
We write
$$
\mathcal{S}= \bigcup\limits_{j\geq 1} \, S_j,
$$
where $S_j$ are the subintervals of $\mathcal{S}$ ordered in an increasing manner. By Lemma \ref{lem-pre1}, we have $S_j = (E_j^-,  E_j^+)$, and it follows that  
$0=E_1^- < E_1^+ \leq E_2^- < E_2^+ \leq  E_3^- < E_3^+ \cdots$. 

(2) follows from Lemmas \ref{lem-eigen1} and \ref{lem-detE}. The eigenvalues of $M(E)$ are given by $\lambda_{E, 1}=e^{ik}$ and $\lambda_{E, 2}=e^{-ik}$.

(3) follows from (1) in Lemma \ref{lem-pre1}. 

(4) From (3), either $E_j^+ = E_j(0)$ or $E_j^+ =E_j(\pi)$ holds.
We consider the former case and show that $E_{j+1}^- = E_{j+1}(0)$, and the proof for the latter case is similar. 
Indeed, note that $|D(E)| > 2$ on the interval $(E_j^+ , E_{j+1}^-)$. Since $D(E_j^+) =2$, 
it follows that $D(E_{j+1}^-) =2$ by Lemma \ref{lem-12}.  That is when $k=0$ in the equation \eqref{eq-disp} 
and we obtain $E_{j+1}^- = E_{j+1}(0)$.

(5) From (2)-(3) in Lemma \ref{lem-pre1}, we see that $|D(E)|<2$ for $E$ sufficiently close to but not equal to $E^*$. 
Therefore, $E^*$ separates two subintervals in $\mathcal{S}$ and consequently $E^*= E_j^+ = E_{j+1}^-$ for some $j\geq 1$. 
Finally, the last assertion follows from the equation \eqref{eq-disp}.  \qed

\subsection{Dirac point}\label{sec-Dirac_point}

A pair $(k^*,E^*)\in\B\times\R$ on the dispersion curves is called a Dirac point if 
\begin{enumerate}
\item [(1)] There exits integer $j\ge1$ such that $E_j(k^*) = E_{j+1}(k^*)=E^*$. In addition, there exit constants 
$\alpha>0$ and $\delta>0$ such that the following expansions
\begin{eqnarray*}
E_j(k)&=&E^* - \alpha |k-k^*| + O((k-k^*)^2),   \\
E_{j+1}(k)&=&E^* + \alpha |k-k^*|+ O((k-k^*)^2)
\end{eqnarray*}
hold for $|k-k^*| < \delta$.
\item [(2)] The multiplicity of the Bloch modes in $L_{k^*}^2$ for the eigenvalue $E^*$ is two.
\end{enumerate}

By virtue of Theorem \ref{thm-bandstructure}, Dirac points can only occur when $k^*=0$ or $k^*=\pi$ with $D(E^*)= \pm2$ and $D'(E^*)=0$
so that $E^*= E_j^+ = E_{j+1}^-$ for some $j\geq 1$.
In fact, as shown below, all pairs $(k^*,E^*)$ satisfying these conditions are Dirac points.



\begin{prop}
Let $E^*\in \R$ with $D'(E^*)=0$. The pair $(k^*,E^*)$ is a Dirac point when $k^*=0$ and $D(E^*)=2$, or when $k^*=\pi$ and $D(E^*)=-2$.
In addition, 
\begin{enumerate}
\item [(1)] The eigenvalues $\lambda_{E, 1}$ and $\lambda_{E, 2}$ of $M(E)$ adopt the following expansions:
\begin{eqnarray*}
\lambda_{E, 1} &=& 1+ i|E-E^*|\sqrt{\frac{1}{2}|D''(E^*)|} + O(E-E^*)^2, \\
\lambda_{E, 2} &=& 1- i|E-E^*|\sqrt{\frac{1}{2}|D''(E^*)|} + O(E-E^*)^2;
\end{eqnarray*}

\item [(2)] The dispersion curves $E_j(k)$ and $E_{j+1}(k)$ adopt the expansions:
\begin{eqnarray}
E_j(k) &=& E^* -|k-k^*|\sqrt{\frac{2}{|D''(E^*)|}} + O((k-k^*)^2), \label{eq-dis1}\\
E_{j+1}(k) &=& E^* + |k-k^*|\sqrt{\frac{2}{|D''(E^*)|}} + O((k-k^*)^2), \label{eq-dis2}
\end{eqnarray}

\item [(3)]
$M'(E^*)$ attains two eigenvalues $\pm i \sqrt{\frac{1}{2}|D''(E^*)|}$. Moreover, there exists two $v_1, v_2 \in \R^2$ such that
$$
M'(E) v_1= -\sqrt{\frac{1}{2}|D''(E^*)|} \, v_2, \quad M'(E) v_2= \sqrt{\frac{1}{2}|D''(E^*)|} \, v_1. 
$$
\end{enumerate}
\end{prop}

\noindent \textbf{Proof.} We provide the proof  for $k^*=0$, and the proof for $k^*=\pi$ follow the same lines.
First $E^*= E_j^+ = E_{j+1}^-$ for some $j\geq 1$ by Theorem \ref{thm-bandstructure}.
Note that $M(E^*)=Id$, thus the multiplicity of the eigenvector is $2$. We deduce from 
Lemma \ref{lem-eigen1} that the multiplicity of the Bloch modes is $2$. \\

 \noindent (1). Note that 
$$
D(E^*)=2, D'(E^*)=0, D''(E^*) <0, M(E^*)=Id.
$$
We obtain
\begin{eqnarray*}
D(E) &=& D(E^*) + D'(E^*)(E-E^*) + \frac{1}{2}D''(E^*)(E-E^*)^2+ O(E-E^*)^3 \\
        &=&2+  \frac{1}{2}D''(E^*)(E-E^*)^2+ O(E-E^*)^3.
\end{eqnarray*}
The expansions for $\lambda_{E, 1}$ and $\lambda_{E, 2}$ follow by substituting the above into the expressions
$\lambda_{E, 1}= \frac{D(E) + i \sqrt{4- D(E)^2}}{2}$ and 
$\lambda_{E, 2}= \frac{D(E) - i \sqrt{4- D(E)^2}}{2}$.  \\

 \noindent  (2). In the neighborhood of $k^*=0$, we have $\cos k = 1- \frac{1}{2}k^2 +O(k^4)$. Solving the equation
$$
2 \cos k = D(E) = 2+  \frac{1}{2}D''(E^*)(E-E^*)^2+ O(E-E^*)^3, 
$$
gives the expansions for two dispersion relations $E_j(k)$ and $E_{j+1}(k)$.  \\

 \noindent (3). 
Note that $M(E) = Id + M'(E)(E-E^*) + O(E-E^*)^2$. Using the asymptotic of the eigenvalues $\lambda_{E, l}$ ($j=1, 2$) for $M(E)$ near $E=E^*$, 
it follows that $M'(E^*)$ has two eigenvalues $\pm i \sqrt{\frac{1}{2}|D''(E^*)|}$. 
Let $v=v_1 + iv_2$ be an eigenvector of the eigenvalue $i \sqrt{\frac{1}{2}|D''(E^*)|}$. Since $M'(E)$ is real-valued, $\bar v= v_1 -iv_2$ is an eigenvector of the eigenvalue $i \sqrt{\frac{1}{2}|D''(E^*)|}$. It follows that
$$
M'(E) v_1= -\sqrt{\frac{1}{2}|D''(E^*)|}v_2, \quad M'(E) v_2= \sqrt{\frac{1}{2}|D''(E^*)|}v_1. 
$$

\qed

From above discussions, Dirac point appears when two neighboring bands cross each other and the multiplicity of the Bloch modes at the Dirac point is two.
In fact, the multiplicity of the Bloch modes for non-Dirac point is always one, as stated in following proposition.

\begin{prop} \label{lem-multi}
For each pair of $(k, E)\in \B \times E_j(k)$ ($j\ge1$) that is not a Dirac point, the multiplicity of the Bloch modes at $(k,E)$ is one. 
\end{prop}

\noindent\textbf{Proof.} In light of Lemma \ref{lem-eigen1}, we only need to show that the multiplicity of the eigenvector for $M(E)$ is 1.
The claim is automatically true when $M(E)$ attains two different eigenvalues,
thus it is sufficient to consider $M(E)$ at $k=0$ and $k=\pi$ only when $\lambda_{E, 1} = \lambda_{E, 2}$. 
Without loss of generality, we consider the former case. Let $E=E_j(0)$, then $\lambda_{E, 1} = \lambda_{E, 2}=1$.
 If $dim Ker (M(E)-Id)=2$, then $M(E)=Id$, which implies that $E_j^+ = E_{j+1}^-$ by Lemma \ref{lem-pre1} and Theorem \ref{thm-bandstructure},
 and consequently $(k=0, E=E_j(0))$ would be a Dirac point.
 Therefore, $dim Ker (M(E)-Id)=1$.  \qed

\subsection{Construction of Bloch modes}\label{sec-bloch_states}



\begin{defn}
We call that the $j$-th band in the dispersion relation $E_j(k)$ is isolated if $E_{j-1}^+ < E_{j}^-$ and  $E_j^+ < E_{j+1}^-$. 
\end{defn}






\medskip

Let $E_j(k)$ be an isolated band. 
For each $E\in [E_j^-, E_j^+]$, let $k$ be a real number in  $[0, \pi]$ such that 
\be  \label{eq-E-k}
e^{ik} = \lambda_{E, 1} = \frac{D(E) + i \sqrt{4- D(E)^2}}{2},
\ee
the first eigenvalue of the matrix $M(E)$.
We choose the associated eigenvector $(a_1, a_2)$ with
$$
a_1= \psi_{E, 2}(1), \quad a_2 = e^{ik} - \psi_{E, 1}(1).
$$
It follows from Lemma \ref{lem-eigen1} that
\begin{equation}\label{eq-phi_jk}
\phi_{j,k}(x)= \psi_{E, 2}(1) \psi_{E, 1}(x) + (e^{ik} - \psi_{E, 1}(1))\psi_{E, 2}(x)
\end{equation}
is a Bloch mode and it forms a basis of the one dimensional eigenspace.
We define the normalized Bloch mode by letting
$$
    \varphi_{j,k} = \frac{\phi_{j,k}}{\|\phi_{j,k}\|_{X}}.
$$
It is clear that the above Bloch mode is well-defined as long as $a_1$ and $a_2$ are not zero simultaneously and the function $\phi_{j,k} \not\equiv 0$.
%
The case $\phi_{j,k} \equiv 0$ is a degenerate case which only occurs at the band edge where $k = 0$ or  $\pi$.
We next show that the Bloch mode $\varphi_{j,k}$ constructed above can be extended continuously  in $[0, \pi]$ when such degeneracy is present.

\begin{lem}
If $\phi_{j ,0}\equiv0$ or $\phi_{j ,\pi}\equiv0$ at the $j$-th band edge, then there holds
$$
\lim_{k\to 0^+} \varphi_{j,k} = \frac{i \psi_{E^*, 2}}{\|\psi_{E^*, 2}\|_{X}}, \quad  E^* =E_j(0). \; 
$$
or
$$
\lim_{k\to \pi^-} \varphi_{j,k} = \frac{i \psi_{E^*, 2}}{\|\psi_{E^*, 2}\|_{X}},  \quad \; E^* =E_j(\pi)
$$
respectively.
Moreover, $\psi_{E^*, 2}$ is a Bloch mode for $k=0$ and $k=\pi$ respectively.

\end{lem}
\noindent\textbf{Proof.} If $\phi_{j ,0}\equiv0$, then $\psi_{E^*, 2}(1) = 1 - \psi_{E^*, 1}(1)=0$ with $E^*= E_j(0)$. 
From \eqref{eq-phi_jk} we have 
$$
\frac{ \partial \phi_{j,k}(x)}{\partial k} 
= \frac{ \partial }{\partial E}
\big(\psi_{E, 2}(1) \psi_{E, 1}(x)- \psi_{E, 1}(1)\psi_{E, 2}(x) + e^{ik} \psi_{E, 2}(x) \big)  E_j'(k) 
+ ie^{ik}\psi_{E, 2}(x). $$
Note that at $E=E^*=E_j(0)$, we have $E_j'(0)=0$. Thus
$$
\frac{ \partial \phi_{j,k}(x)}{\partial k}|_{k=0} = i\psi_{E, 2}(x).
$$
It follows that
$$
\varphi_{j,0^+} = \lim_{k\to 0^+} \varphi_{j,k} = \frac{i \psi_{E, 2}}{\|\psi_{E, 2}\|_{X}}.
$$ 
We now show that $\psi_{E^*, 2}$ is a Bloch mode. Indeed, since 
$\psi_{E^*, 2}(1) = 0$ and  $\psi_{E^*, 1}(1)=1$, from $\eqref{eq-E-k}$ we have $\frac{1}{\mu(0)}\psi_{E^*, 2}'(1)=1 =\frac{1}{\mu(0)}\psi_{E^*, 2}'(0) $. 
Therefore, it follows from Lemma \ref{lem-eigen1} that $\psi_{E^*, 2}$ is a Bloch mode for $k=0$.  
The proof for the other case is similar. \qed

\medskip

From the above discussions, we see that the Bloch modes
\be
\varphi_{j,k} = 
\begin{cases}
\frac{ \phi_{j,k}}{\|\phi_{j,k}\|_{X}},  \mbox{ if } \phi_{j, k} \not\equiv 0, \\
\frac{i \psi_{E, 2}}{\|\psi_{E, 2}\|_{X}}, \mbox{ if } \,\, \phi_{j,k} \equiv 0
\end{cases} 
\ee
are continuous in the reduced Brillouin zone $[0, \pi]$.
They can be extended for $k\in (-\pi, 0)$ by letting
$$
\varphi_{j,k} = \bar \varphi_{j, -k} = \frac{\bar \phi_{j,-k}}{\|\phi_{j,-k}\|_{X}}.
$$
%
%
In summary, we have constructed the Bloch modes in the $j$-th isolated band as follows.
\begin{prop}
Let $\phi_{j,k}$ be defined by \eqref{eq-phi_jk}. Then
the Bloch modes over the $j$-th band 
\be  \label{eq-gauge1}
\varphi_{j,k} = 
\begin{cases}
\frac{ \phi_{j,k}}{\|\phi_{j,k}\|_{X}},  0\leq  k \leq  \pi \; \mbox{and} \; \phi_{j,k} \not\equiv 0\\
\frac{i \psi_{E, 2}}{\|\psi_{E, 2}\|_{X}},  \,\, k\in \{0, \pi\}  \; \mbox{and} \; \phi_{j,k} \equiv 0,\\
\bar \varphi_{j,-k}, \quad -\pi < k <0 
\end{cases}
\ee
is smooth for $k \in ( -\pi, 0)\cup (0, \pi)$.  
In the degenerate case when $\phi_{j,k}(x)\equiv0$, there holds
$$
\varphi_{j,0^-} = \lim_{k\to 0^-} \varphi_{j,k} =  \bar\varphi_{j,0}=\frac{-i \psi_{E, 2}}{\|\psi_{E, 2}\|_{X}} \quad \mbox{and} 
\quad \varphi_{j,(-\pi)^+} = \lim_{k\to (-\pi)^+} \varphi_{j,k} =  \bar\varphi_{j,\pi}=\frac{-i \psi_{E, 2}}{\|\psi_{E, 2}\|_{X}}.
$$

\end{prop}


\subsection{Zak phase}\label{sec-Zak} 
\subsubsection{Zak phase for an isolated band}\label{sec-Zak1} 

For a given normalized Bloch mode $\varphi_{j,k}$, one can express $\varphi_{j,k}$ in the form of
$$
\varphi_{j,k}(x) = e^{ikx}u_{j, k}(x), 
$$
where $u_{j, k}(x)$ is a periodic function satisfying $u_{j, k}(x)=u_{j, k}(x+1)$. 
$u_{j, k}$ is called the periodic part of the Bloch mode $\varphi_{j,k}$. 
For the $j$-th band that is isolated, 
it is clear that the Bloch modes $\varphi_{j,k}$ 
can form a closed loop as $k$ runs over the Brillouin zone from $-\pi$ to $\pi$ since $\varphi_{j,\pi}$ and $\varphi_{j,-\pi}$ only differs by a global phase constant.
However, this is no longer the case for the periodic part 
$u_{j, k}$ since $u_{j, \pi}(x)=e^{-i2\pi x}u_{j, -\pi}(x)$ even if $\varphi_{j,\pi} = \varphi_{j,-\pi}$. To take this into account, we
define the following discrete Zak phase over the $j$-th band (cf. Section 3.4 in \cite{Vanderbilt-18})
\begin{eqnarray}
 && \theta_j^{(N)} = \sum_{n=1}^{N-1}  -\Im \ln \left(u_{j, k_{n+1}} , u_{j, k_n} \right)_{X} -\Im \ln \left(e^{-i2\pi x} u_{j, k_0} , u_{j, k_{N-1}} \right)_{X} \,\, mod \,\, 2\pi, \label{eq-zak_discrete} \\
&&\mbox{where} \;  k_n = -\pi+\frac{2\pi n}{N}. \nonumber
\end{eqnarray}
 If the Bloch mode $\varphi_{j,k}$ is smooth with respect to $k$ over the Brillouin zone with $\varphi_{j,-\pi}=\varphi_{j,\pi}$, 
by taking the continuum limit of \eqref{eq-zak_discrete} as $N\to\infty$, we recover the well-known Zak phase formula  (cf. \cite{Vanderbilt-18, Zak-89})
 \begin{equation}\label{eq-zak_int}
 \theta_j = i \int_{-\pi}^{\pi} \left(\frac{\partial u_{j,k}}{\partial k}, u_{j, k} \right)_{X} \, dk \,\, mod \,\, 2\pi.    
 \end{equation}
The Zak phase $\theta_j$ is invariant with respect to the Gauge transformation.
This can be observed from the discrete formulation \eqref{eq-zak_discrete}. On the other hand, it is attempting to use the Bloch mode $\varphi_{j,k}$ instead of its periodic part $u_{j, k}$ in formula (\ref{eq-zak_int}) since the closed-loop property enjoyed by the former, however, a Zak phase defined as such will depend on the choice of the periodic cell in the photonic structure by a straightforward calculation. 
 
For the Bloch modes $\varphi_{j,k}$ constructed in Section \ref{sec-bloch_states}, the periodic part $u_{j, k}$ is piecewisely smooth in $(-\pi, 0)\cup (0, \pi)$ with respect to $k$. Thus the continuous formula \eqref{eq-zak_int} can not be used directly. We revise \eqref{eq-zak_int} by taking into account of the possible phase jump at  $k=0, \pi$, and define the Zak phase accordingly as
\begin{equation} \label{eq-zak1}
\theta_j =  \; i \int_{-\pi}^{0}\left(\frac{\partial u_{j,k}}{\partial k}, u_{j, k}\right)_{X} \, dk +  i \int_{0}^{\pi}\left(\frac{\partial u_{j,k}}{\partial k}, u_{j, k}\right)_{X} \, dk + \hat\theta_j \quad mod\,\, 2\pi,
\end{equation}
where $\hat\theta_j$ is given by
\begin{equation*}
\hat\theta_j = -\Im \ln (u_{j, 0}, u_{j, 0^-})_X - \Im \ln ( e^{-i2\pi x} u_{j, (-\pi)^+},u_{j, \pi})_X. 
\end{equation*}
In the above, $u_{j, 0^-} = \lim_{k\to 0^-} u_{j, k}$ and $u_{j, (-\pi)^+} = \lim_{k\to (-\pi)^+} u_{j, k}$  denote the one-side limit.

If $\phi_{j,0}$ and $\phi_{j,\pi}$ are non-degenerate, noting that $\phi_{j, 0}$ and $\phi_{j, \pi}$ are real-valued,
it follows that
$$
-\Im \ln (u_{j, 0}, u_{j, 0^-})_X = 0  \quad \mbox{and} \quad -\Im \ln ( e^{-i2\pi x} u_{j, (-\pi)^+},u_{j, \pi})_X = 0
$$
Otherwise, using the construction in \eqref{eq-gauge1}, we have
$$
-\Im \ln (u_{j, 0}, u_{j, 0^-})_X = \pi \,\, mod \,\, 2\pi  \quad \mbox{and} \quad -\Im \ln ( e^{-i2\pi x} u_{j, (-\pi)^+},u_{j, \pi})_X = \pi \,\, mod \,\, 2\pi
$$
respectively. On the other hand, there holds
$$
\int_{-\pi}^{0}\left(\frac{\partial u_{j,k}}{\partial k}, u_{j, k}\right)_{X} \, dk   
= \int_{0}^{\pi} \left(-\frac{\partial \bar u_{j,k}}{\partial k}, \bar u_{j, k} \right )_{X} \, dk 
= - \overline{ \int_{0}^{\pi} \left(\frac{\partial u_{j,k}}{\partial k},  u_{j, k}\right)_{X} \, dk }.
$$


\noindent In summary, we obtain the following formula for the Zak phase.
\begin{prop}
The Zak phase for an isolated band $E_j(k)$ is 
\be
\theta_j = 2 \Im \int_{0}^{\pi}\left(\frac{\partial u_{j,k}}{\partial k}, u_{j, k}\right)_{X} \, dk + 
\begin{cases}
\pi, \quad \mbox{if } \,\, \phi_{j, 0}\equiv0 \not\equiv \phi_{j, \pi}, \mbox{or} \,\, \phi_{j, \pi}\equiv0 \not\equiv \phi_{j, 0}, \\
0, \quad \mbox{otherwise }.
\end{cases}
\ee
\end{prop}

\subsubsection{Zak phase at the presence of Dirac point}\label{sec-Zak2}
Assume that the band $E_j(k)$ and $E_{j+1}(k)$ cross at the Dirac point $(0, E^*)$.
Let
$$
\phi_{j,k}(x)= \psi_{E, 2}(1) \psi_{E, 1}(x) + (e^{ik} - \psi_{E, 1}(1))\psi_{E, 2}(x),
$$
where $E=E_j(k)$ is given by (\ref{eq-dis1}). 
Then we have the following expansions for $k>0$ and $E< E^*$:
\begin{align}
e^{ik} & = 1+ ik + O(k^2) = 1- i(E-E^*)\sqrt{-\frac{1}{2}D''(E^*)}+O(E-E^*)^2, \\
\psi_{E, 2}(1) & = \frac{\partial \psi_{E, 2}(1)}{\partial E}(E^*) (E-E^*)+  O(E-E^*)^2, \\
\psi_{E, 1}(1) & = 1+ \frac{\partial \psi_{E, 1}(1)}{\partial E}(E^*) (E-E^*)+  O(E-E^*)^2.  
\end{align}

Define
$$
b_1(E) = \frac{\partial \psi_{E, 1}(1)}{\partial E}, \quad b_2(E) = \frac{\partial \psi_{E,2}(1)}{\partial E}. 
$$
we see that $\phi_{j,k}$ adopts the expansion
$$
\phi_{j,k}(x)= (E-E^*) \left( b_2(E^*)\psi_{E, 1}(x)  + \left( -i \sqrt{-\frac{1}{2}D''(E^*)}-  b_1(E^*)\right)\psi_{E, 2}(x)\right) + O (E-E^*)^2.
$$
It follows that
$$
\varphi_{j,0}(x) = -\frac{b_2(E^*)\psi_{E^*, 1}(x)  + \left( -i \sqrt{-\frac{1}{2}D''(E^*)}- b_1(E^*) \right)\psi_{E^*, 2}(x)}{\| b_2(E^*)\psi_{E^*, 1}  + \left( -i \sqrt{-\frac{1}{2}D''(E^*)}-  b_1(E^*)\right)\psi_{E^*, 2}\|_X}, 
$$
and
$$
\varphi_{j,k}(x) = \varphi_{j,0}(x) + O(k) \quad  \mbox{for} \quad k>0.
$$
Similarly, there holds
\begin{equation}\label{eq-psi_relation}
\varphi_{j+1,0}(x) = \frac{b_2(E^*)\psi_{E^*, 1}(x)  + \left( i \sqrt{-\frac{1}{2}D''(E^*)}- b_1(E^*)\right)\psi_{E^*, 2}(x)}{\| b_2(E^*)\psi_{E^*, 1}  
+ \left( i \sqrt{-\frac{1}{2}D''(E^*)}-  b_1(E^*)\right)\psi_{E^*, 2}\|_X} = -\overline{\varphi_{j,0}(x)}, 
\end{equation}
and
$$
\varphi_{j+1,k}(x) = \varphi_{j+1,0}(x) + O(k) \quad  \mbox{for} \quad k>0.
$$

Now the Zak phase for the $j$-th band $E_j(k)$ and the $j+1$-th band $E_{j+1}(k)$ is
\begin{align*}
\theta_j &= 2 \Im \int_{0}^{\pi}\left(\frac{\partial u_{j,k}}{\partial k}, u_{j, k}\right)_{X} \, dk - \Im \ln (u_{j, 0}, u_{j, 0^-})_X - \Im \ln (  e^{-i2\pi x} u_{j, (-\pi)^+},u_{j, \pi})_X  \,\, mod\,\, 2\pi  \\
& = 2 \Im \int_{0}^{\pi}\left(\frac{\partial u_{j,k}}{\partial k}, u_{j, k}\right)_{X} \, dk - \Im\ln (\varphi_{j,0},\varphi_{j,0^-})_X - \Im \ln (\varphi_{j,(-\pi)^+},\varphi_{j,\pi})_X  \,\, mod\,\, 2\pi \\
& = 2 \Im \int_{0}^{\pi}\left(\frac{\partial u_{j,k}}{\partial k}, u_{j, k}\right)_{X} \, dk - \Im \ln (\varphi_{j,0},\overline{\varphi_{j,0}})_X - \Im \ln (\overline{\varphi_{j,\pi}},\varphi_{j,\pi})_X  \,\, mod\,\, 2\pi  \\
\theta_{j+1} &= 2 \Im \int_{0}^{\pi}\left(\frac{\partial u_{j+1,k}}{\partial k}, u_{j+1, k}\right)_{X} \, dk - \Im \ln (\varphi_{j+1,0}, \overline{\varphi_{j+1,0}})_X - \Im \ln (\overline{\varphi_{j+1,\pi}},\varphi_{j+1,\pi})_X  \,\, mod\,\, 2\pi.
\end{align*}
By virtue of the relation \eqref{eq-psi_relation}, the following formula holds for the Zak phase over these two bands.
\begin{prop}
Assume that $E_j(k)$ and $E_{j+1}(k)$ cross at the Dirac point $(0, E^*)$ and that they do not cross with other bands in the spectrum,
then 
\be
\theta_j + \theta_{j+1} = 2  \sum_{\ell=j}^{\ell=j+1}  \Im  \int_{0}^{\pi} \left( \frac{\partial u_{\ell,k}}{\partial k}, u_{\ell, k}\right)_{X} \, dk
+ \begin{cases}
\pi, \quad \mbox{if } \,\, \phi_{j, \pi} \equiv0 \not\equiv \phi_{j+1, \pi}, \\
\pi, \quad \mbox{if }  \,\, \phi_{j+1, \pi} \equiv0 \not\equiv \phi_{j, \pi}, \\
0, \quad \mbox{otherwise}.
\end{cases}
\ee
\end{prop}

If $E_j(k)$ and $E_{j+1}(k)$ cross at the Dirac point $(\pi, E_j^+)$,
the parallel lines above lead to the following proposition for the Zak phase.
\begin{prop}
If $E_j(k)$ and $E_{j+1}(k)$ cross at the Dirac point $(\pi, E^*)$, then
\be
\theta_j + \theta_{j+1} = 2  \sum_{\ell=j}^{\ell=j+1}  \Im  \int_{0}^{\pi} \left( \frac{\partial u_{\ell,k}}{\partial k}, u_{\ell, k}\right)_{X} \, dk
+ \begin{cases}
\pi, \quad \mbox{if } \,\, \phi_{j, 0} \equiv0 \not\equiv \phi_{j+1, 0}, \\
\pi, \quad \mbox{if }  \,\, \phi_{j+1, 0} \equiv0 \not\equiv \phi_{j, 0}, \\
0, \quad \mbox{otherwise}.
\end{cases}
\ee
\end{prop}

\subsection{Mode decomposition in the band gap} \label{sec-gapmode}
If $E_j^+ < E_{j+1}^-$, the spectrum of $\Lo$ attains a band gap  $(E_j^+, E_{j+1}^-)$.
The wave fields with frequency in the gap can be decomposed into two modes, one increases exponentially and the other decreases exponentially as $\abs{x}$ increases. 

\begin{rem}
Here and henceforth, without loss of generality we assume that the trace $D(E)>0$ so that
the eigenvalues defined in \eqref{eq-lambda1_lambda2} satisfy
$|\lambda_{E, 1}| <1$ and $|\lambda_{E, 2}| >1$ in the band gap.
If $D(E)<0$, all the arguments follow by replacing $\lambda_{E, 1}$ and $\lambda_{E, 2}$ with each other.
\end{rem}
Assume that $\psi_{E, 2}(1) \neq 0$. We let 
\begin{equation}\label{eq-eigenvec}
 V_{E,1} =  \begin{pmatrix}
 \psi_{E, 2}(1)\\
 \lambda_{E, 1} - \psi_{E, 1}(1)
 \end{pmatrix}  \quad \mbox{and} \quad
 V_{E,2} = \begin{pmatrix}
 \psi_{E, 2}(1)\\
 \lambda_{E, 2} - \psi_{E, 1}(1)
 \end{pmatrix}
\end{equation}
be the eigenvectors of $M(E)$ corresponding to the eigenvalue $\lambda_{E, 1}$ and $\lambda_{E, 2}$, respectively.

Let $u\in L^1_{loc}(\mathbf{R})$ be a solution of $(\Lo -E)u=0$ for $E \in (E_j^+, E_{j+1}^-)$. Define the vector-valued function $U(x):=(u(x), \frac{1}{\mu(x)}u'(x))^T$. Using the transfer matrix $M(E)$, there holds $U(n)=M(E)^n U(0)$.
If one decomposes $U(0)$ as $U(0)=t_1 V_{E,1} + t_2 V_{E,2}$, then it follows that
$$ U(n)= t_1^n V_{E,1} + t_2^n V_{E,2}. $$
This leads to the following lemma.
\begin{lem} \label{lem-gapmode}
Let $E \in (E_j^+, E_{j+1}^-)$ and $\psi_{E, 2}(1) \neq 0$. Let $u\in L^1_{loc}(\mathbf{R})$ be a solution to $\Lo u = Eu$.
\begin{enumerate}
    \item $u(x) \to 0$ as $x \to +\infty$ if and only if $U(0)=c V_{E,1}$ for some nonzero constant $c$. Meanwhile $|u(x)| \to \infty$ as $x \to -\infty$.
    \item $u(x) \to 0$ as $x \to -\infty$ if and only if $U(0)=c V_{E,2}$ for some nonzero constant $c$. Meanwhile, $|u(x)| \to +\infty$ as $x \to +\infty$.
\end{enumerate}
\end{lem}


We now consider the case when $\psi_{E, 2}(1) = 0$. Recall the transfer matrix
$$
M(E)= 
\begin{pmatrix}
\psi_{E, 1}(1) &  0 \\
\frac{1}{\mu(0)}\psi_{E, 1}'(1) &   \frac{1}{\mu(0)} \psi_{E, 2}'(1)
\end{pmatrix}.
$$
We may have $|\psi_{E, 1}(1)|<1$ or $|\psi_{E, 1}(1)|>1$. Let us first consider the case when $|\psi_{E, 1}(1)|<1$. It is clear that the two eigenvalues of 
$M(E)$ are given by $\psi_{E, 1}(1)$ and $\frac{1}{\mu(0)} \psi_{E, 2}'(1)$.
We have 
$\lambda_{E, 1} = \psi_{E, 1}(1)$ and $\lambda_{E, 2} = \frac{1}{\mu(0)} \psi_{E, 2}'(1)$. 
It follows that
$$
|\psi_{E, 1}(x)| \to 0 \quad \mbox{as} \; x \to +\infty
$$ 
while
$$
|\psi_{E, 2}(x)| \to \infty \quad \mbox{as} \; x \to +\infty
$$
Similarly, if $|\psi_{E, 1}(1)|>1$, we have $\lambda_{E, 2} = \psi_{E, 1}(1)$ and $\lambda_{E, 1} = \frac{1}{\mu(0)} \psi_{E, 2}'(1)$. Moreover,
$$
|\psi_{E, 1}(x)| \to \infty \quad \mbox{as} \,\, x \to +\infty
$$ 
while
$$
|\psi_{E, 2}(x)| \to 0 \quad \mbox{as} \,\, x \to +\infty. 
$$

\begin{lem}
Assume that the $j$-th band gap is open and $E \in (E_j^+, E_{j+1}^-)$. If 
$\psi_{E, 2}(1) = 0$, then we have the following two cases:
\begin{enumerate}
\item
Case A. $|\psi_{E, 1}(1)|<1$: $\psi_{E, 1}$ and $\psi_{E, 2}$ is an interface mode for the following semi-infinite system with Neumann and Dirichlet boundary condition respectively:
$$
\begin{cases}
(\Lo -E) \psi_{E, 1}=0, \quad x>0, \\
\psi_{E, 1}'(0)=0. 
\end{cases}
\quad
\begin{cases}
(\Lo -E) \psi_{E, 2}=0, \quad x<0, \\
\psi_{E, 2}(0)=0. 
\end{cases}
$$
$$
$$
Moreover, if $u$ is a solution to $\Lo u = Eu$. Then there holds
$\lim_{x \to +\infty} u(x) =0$ and  $\lim_{x \to -\infty} u(x) =0$ respectively if and only if $u= c\psi_{E, 1}$  and $u= c\psi_{E, 2}$ respectively for some nonzero constant $c$.

\item
Case B. $|\psi_{E, 1}(1)|>1$: 
$\psi_{E, 2}$ and $\psi_{E, 1}$ is an interface mode for the following semi-infinite system with Neumann and Dirichlet boundary condition respectively:
$$
\begin{cases}
(\Lo -E) \psi_{E, 2}=0, \quad x>0, \\
\psi_{E, 2}'(0)=0. 
\end{cases}
\quad
\begin{cases}
(\Lo -E) \psi_{E, 1}=0, \quad x<0, \\
\psi_{E, 1}(0)=0. 
\end{cases}
$$
Moreover, if $u$ is a solution to $\Lo u = Eu$. Then there holds
$\lim_{x \to +\infty} u(x) =0$ and  $\lim_{x \to -\infty} u(x) =0$ respectively if and only if $u= c\psi_{E, 2}$ and $u= c\psi_{E, 1}$ respectively for some nonzero constant $c$.


\item
Similar conclusions also hold if $\psi_{E, 1}'(1)=0$ or $\lambda_{E, 1} -\psi_{E, 1}(1) =0$. 
\end{enumerate}
\end{lem}

Finally, we present a lemma that will be used in the proof of Lemma \ref{lem-edgemode2}. 
\begin{lem} \label{lem-e2}
If $\psi_{E, 2}(1)=0$, then $\psi_{E, 2}'(1) \neq 0$ and
$$
\psi_{E, 2}'(1)\cdot \frac{\partial \psi_{E, 2}(1)}{\partial E} = \mu(0) \int_{0}^1 \psi_{E, 2}^2(x)\varepsilon(x) dx >0.
$$
\end{lem}
\noindent\textbf{Proof.} 
Note that  
$$
\begin{cases}
(\Lo-E)\psi_{E, 2} = 0, \\
\psi_{E, 2}(0) = 0, \frac{1}{\mu(0)} \psi_{E, 2}'(1) =1.
\end{cases}
$$
Let $\psi (x) = \frac{\partial \psi_{E, 2}(x)}{\partial E }$, then
$$
\begin{cases}
(\Lo-E)\psi = \psi_{E, 2} , \\
\psi(0) = 0,  \psi'(1) =0.
\end{cases}
$$
Therefore
$$
\int_{0}^1 \left( (\frac{1}{\mu} \psi')' + E\varepsilon(x) \psi  \right) \psi_{E, 2}(x) dx = - \int_{0}^1 \psi_{E, 2}^2(x)\varepsilon(x) dx. 
$$
Integration by part and using the boundary conditions that 
$\psi(0) = 0,  \psi'(1) =0, \psi_{E, 2}(0)=0$
yield
$$
\frac{1}{\mu(1)} \psi_{E, 2}'(1) \psi(1) = \int_{0}^1 \psi_{E, 2}^2(x)\varepsilon(x) dx >0,
$$
which yields the desired estimate.  \qed

\section{Interface modes induced by Dirac points for general time-reversal symmetric structures }\label{sec:perturb_sys}
In this section, we study the perturbation to a general time-reversal symmetric photonic structure with a Dirac point.
Assume that the operator $\Lo$ for the photonic structure with parameters $(\mu(x), \varepsilon(x))$ attains a Dirac point $(k^*,E^*)$ at the intersection of the $j$-th and the $j+1$-th band.
Without loss of generality, we consider the Dirac point with $k^*=0$ and 
$$E^*= E_j^+ = E_{j+1}^-= E_j(0)=E_{j+1}(0).$$ 
We shall derive conditions for the perturbation such that a band gap opens near the Dirac point. 
Furthermore, we derive conditions on the perturbation which guarantee the existence of an interface mode near the interface which separates two periodic structures after the perturbation. 

\subsection{Perturbation of photonic systems with Dirac points}

We perturb the photonic system associated with the operator $\Lo$ in the following way:
$$
\begin{cases}
\mu(x)  \to \mu(x) + \delta \tilde \mu(x),\\
\varepsilon(x) \to \varepsilon(x) + \delta \tilde \varepsilon(x),
\end{cases}
$$
where $|\delta| \ll 1$, and $\tilde \mu(x)$ and $\tilde \varepsilon(x)$ are two piecewisely continuous periodic functions (with period 1) 
satisfying $\norm{\tilde \mu}_{L^{\infty}}+\norm{\tilde \varepsilon}_{L^{\infty}}=1$.  
The perturbed operator is denoted by
$$
 \Lo_{\delta}\psi(x) = -\frac{1}{\varepsilon(x) + \delta \tilde \varepsilon(x)}\left( \frac{1}{\mu(x) + \delta \tilde \mu(x)}\psi'(x)\right)'.
$$
For each $E\in \mathbf{R}$, 
let $\psi_{E,1, \delta}$ and $\psi_{E, 2, \delta}$ be the unique solution to the following equations respectively:
\begin{align*}
(\Lo_{\delta} - E)   \psi_{E,1, \delta} =0, \quad \psi_{E,1, \delta}(0)= 1, \,\, \frac{1}{\mu(0)+\delta \tilde \mu(0)}\psi_{E,1, \delta}'(0)=0, \\
(\Lo_{\delta}-E)   \psi_{E,2, \delta} =0, \quad \psi_{E,2, \delta}(0)= 0, \,\, \frac{1}{\mu(0)+ \delta \tilde \mu(0)}\psi_{E,2, \delta }'(0)=1.  
\end{align*}
Let $\Psi_{E, \delta}$  denote the perturbed transfer matrix which solves the ODE system
\be
\frac{d}{dx}\Psi_{E, \delta} (x) = J(B + EW + \delta F)\Psi_{E, \delta}(x), \quad  \Psi_{E, \delta}(0)= Id, 
\ee
where 
\begin{equation}\label{eq-F1}
F= 
 \begin{pmatrix}
 E\tilde \varepsilon(x) &  \\
  & \tilde \mu(x)\\
\end{pmatrix}.
\end{equation}
Let $M(E, \delta)=\Psi_{E, \delta} (1)$ be the transfer matrix for one period, and let its two eigenvalues be $\lambda_{E, 1, \delta}$ and $\lambda_{E, 2, \delta}$. 
The trace of $M(E, \delta)$ is denoted as $D(E, \delta)$.

A standard perturbation theory (see \cite{brown-2013} for instance) yields
\begin{align}
\frac{\partial \Psi_{E, \delta} (x)}{\partial E} 
&= \Psi_{E, \delta} (x) \int_{0}^x \Psi_{E, \delta}^{-1}(t)JW(t) \Psi_{E, \delta} (t)dt, \label{eq-perturb1} \\
\frac{\partial \Psi_{E, \delta} (x)}{\partial \delta} 
&= \Psi_{E, \delta} (x) \int_{0}^x \Psi_{E, \delta}^{-1}(t)JF(t) \Psi_{E, \delta} (t)dt, \label{eq-perturb2}\end{align}
and
\begin{align}
\frac{\partial^2 \Psi_{E, \delta} (x)}{\partial E^2} 
&= 2\Psi_{E, \delta} (x) \int_{0}^x \Psi_{E, \delta}^{-1} (t)JW(t) \frac{\partial \Psi_{E, \delta} (t)}{\partial E}dt, \label{eq-perturb3}\\
\frac{\partial^2 \Psi_{E, \delta} (x)}{\partial \delta^2} 
&=2\Psi_{E, \delta} (x) \int_{0}^x \Psi_{E, \delta}^{-1} (t)JF(t) \frac{\partial \Psi_{E, \delta} (t)}{\partial \delta}dt,\label{eq-perturb4} \\
\frac{\partial^2 \Psi_{E, \delta} (x)}{\partial E \partial \delta} 
&= \Psi_{E, \delta} (x) \int_{0}^x \Psi_{E, \delta}^{-1} (t)J\left( W(t) \frac{\partial \Psi_{E, \delta} (t)}{\partial \delta}+ F(t) \frac{\partial \Psi_{E, \delta} (t)}{\partial E}\right)dt. \label{eq-perturb5}
\end{align}

Note that $\Psi_{E^*, 0} (1)=Id,$ $D(E^*, 0)=2$. For simplicity of notation, we write
$$
\begin{cases}
Q_1(x) = \Psi_{E^*, 0}^{-1} (x) J W(x)\Psi_{E^*, 0} (x), \\
Q_2(x) = \Psi_{E^*, 0}^{-1} (x) J F(x)\Psi_{E^*, 0} (x),
\end{cases}
$$
We also write $\Psi_{E^*, 0} (x)  = (u(x), v(x))$, with 
\begin{eqnarray}\label{eq-uv}
u= 
\begin{pmatrix}
\psi_{E^*, 1, 0}(x)\\
\frac{1}{\mu(0)}\psi_{E^*, 1, 0}'(x)
\end{pmatrix}, 
\quad
v= 
\begin{pmatrix}
\psi_{E^*, 2, 0}(x)\\
\frac{1}{\mu(0)}\psi_{E^*, 2, 0}'(x)
\end{pmatrix}.
\end{eqnarray}
Using the fact that $det \, \Psi_{E^*, 0}  =1$, we have
$$
\Psi_{E^*, 0}^{-1} (x) = \begin{pmatrix}
\frac{1}{\mu(0)}\psi_{E^*, 2, 0}'(x) & -\psi_{E^*, 2, 0}(x)\\
-\frac{1}{\mu(0)}\psi_{E^*, 1, 0}'(x) & \psi_{E^*, 1, 0}(x)
\end{pmatrix}.
$$
It follows from a direct calculation that
\be \label{eq-Q}
Q_1= \begin{pmatrix}
-v^TWu & -v^T W v \\
u^TWu & u^TWv
\end{pmatrix}, \quad
Q_2= \begin{pmatrix}
-v^TFu & -v^T F v \\
u^TFu & u^TFv
\end{pmatrix}.
\ee

\begin{lem}
The following hold for the derivatives of $D(E,\delta)$:
\begin{eqnarray*}
\frac{\partial D}{\partial E}(E^*, 0) =0, && \frac{\partial D}{\partial \delta}(E^*, 0) =0. \\
 \frac{1}{2}\frac{\partial^2 D}{\partial E^2}(E^*, 0) &=& \left(\int_{0}^1u^TWv dx\right)^2 -
\left(\int_{0}^1v^TWv dx \right)\cdot \left(\int_{0}^1u^TWu dx \right)\\
\frac{1}{2}\frac{\partial^2 D}{\partial \delta^2}(E^*, 0) &=& \left(\int_{0}^1u^TFv dx\right)^2 -
\left(\int_{0}^1v^TFv dx \right)\cdot \left(\int_{0}^1u^TFu dx \right)\\
\frac{1}{2}\frac{\partial^2 D}{\partial E \partial \delta}(E^*, 0) &=& \left(\int_{0}^1u^TWv dx\right)\cdot \left(\int_{0}^1u^TFv dx\right) 
-\frac{1}{2}\left(\int_{0}^1v^TWv dx \right)\cdot \left(\int_{0}^1u^TWu dx\right) \nonumber\\
&&  -\frac{1}{2}\left(\int_{0}^1v^TFv dx \right)\cdot \left(\int_{0}^1u^TFu dx\right).
\end{eqnarray*}

\end{lem}
\textbf{Proof.} Let $x=1$ in (\ref{eq-perturb1}). By noting that $\Psi_{E^*, 0} (1)=Id$, we have
$$
\frac{\partial M}{\partial E}(E^*, 0) 
= \int_{0}^1 \Psi_{E^*, 0}^{-1}(t)JW(t) \Psi_{E^*, 0} (t)dt.  
$$
Taking the trace and using the fact that $Tr \,AB = Tr\, BA$, and that $Tr \, JW(t)=0$, 
we obtain
$$
\frac{\partial D}{\partial E}(E^*, 0) = \int_{0}^1 Tr\, JW(t) dt =0. 
$$
$\frac{\partial D}{\partial \delta}(E^*, 0) =0$ follows similarly by using \eqref{eq-perturb2}.

We next show that
\begin{align}
\frac{\partial^2 D}{\partial E^2}(E^*, 0) &= Tr \, \left[\int_{0}^1 Q_1(x)dx \right]^2, \label{eq-second1}\\
\frac{\partial^2 D}{\partial \delta^2}(E^*, 0) &= Tr \, \left[\int_{0}^1 Q_2(x)dx \right]^2,\label{eq-second2}\\
\frac{\partial^2 D}{\partial E \partial \delta}(E^*, 0) &= Tr \, \left[\int_{0}^1 Q_1(x)dx \right] \left[\int_{0}^1 Q_2(x)dx\right]. \label{eq-second3}
\end{align}

In light of (\ref{eq-perturb3}), we have
\begin{align*}
\frac{\partial^2 M}{\partial E^2} (E^*, 0)
&= 2\int_{0}^1 \Psi_{E, 0}^{-1} (x)JW(x) \frac{\partial \Psi_{E, 0} (x)}{\partial E}dx \\
&= 2\int_{0}^1 \Psi_{E^*, 0}^{-1} (x)JW(x) \Psi_{E^*, 0} (x) \int_{0}^x \Psi_{E^*, 0}^{-1}(t)JW(t) \Psi_{E^*, 0} (t)dt dx   \\
&= 2 \int_{0}^1 \int_{0}^xQ_1(x)Q_1(t)dtdx.   
\end{align*}
Taking the trace, we get 
\begin{align*}
\frac{\partial^2 D}{\partial E^2}(E^*, 0) &= 
2\int_{0}^1 \int_{0}^x Tr \, Q_1(x)Q_1(t)dtdx \\ 
&=\int_{0}^1 \int_{0}^x Tr \, Q_1(x)Q_1(t)dtdx + \int_{0}^1 \int_{0}^t Tr \, Q_1(t)Q_1(x)dxdt \\
&= \int_{0}^1 \int_{0}^x Tr \, Q_1(x)Q_1(t)dtdx + \int_{0}^1 \int_{x}^1 Tr \, Q_1(x)Q_1(t)dtdx
&= \int_{0}^1 \int_{0}^1 Tr \, Q_1(x)Q_1(t)dtdx \\
&= Tr \, \int_{0}^1 \int_{0}^1 Q_1(x)Q_1(t)dtdx =  Tr \, (\int_{0}^1  Q_1(x)dx)^2.
\end{align*}
which proves \eqref{eq-second1}. The equality (\ref{eq-second2})-(\ref{eq-second3}) can be proved in a similar manner. Finally, the desired results follow from (\ref{eq-Q}) and a direct calculation. \qed

%

\subsection{Band gap opening for the perturbed system}

Let us denote
$$
a_1 = \frac{\partial^2 D}{\partial E^2}(E^*, 0), \quad a_2= \frac{\partial^2 D}{\partial E \partial \delta}(E^*, 0), \quad 
a_3 = \frac{\partial^2 D}{\partial \delta^2}(E^*, 0).
$$

\begin{lem} \label{lem-a1}
There holds
$
a_1 = \frac{\partial^2 D}{\partial E^2}(E^*, 0)<0.
$
\end{lem}

\noindent\textbf{Proof.} Note that 
$$
\int_{0}^1v^TWv = \int_{0}^1 \psi_{E, 2}^2(x)\varepsilon(x) dx >0,\,\, 
\int_{0}^1u^TWu = \int_{0}^1 \psi_{E, 1}^2(x)\varepsilon(x) dx>0,
$$
and that
$$
\int_{0}^1u^TWv = \int_{0}^1 \psi_{E, 2}(x)\psi_{E, 1}(x)\varepsilon(x) dx. 
$$
Since $\psi_{E, 2}$ and $\psi_{E, 1}$ are linearly independent, using Cauchy-Schwartz type inequality, we can derive that
$$
\left|\int_{0}^1u^TWv dx \right|^2 < \int_{0}^1v^TWv dx \cdot \int_{0}^1u^TWu dx, 
$$
whence $a_1 <0$ follows. 

\qed

\begin{thm}\label{thm-bandgap}
Let $\delta>0$ be a sufficiently small number. Assume that
\be \label{eq-assume1}
a_2^2 - a_1 a_3 > 0.
\ee
Then there exists a band gap $(E_{j, \delta}^{+}, E_{j+1, \delta}^{-} )$ 
between the $j$-th and the $(j+1)$-th band for the perturbed operator $\Lo_{\delta}$.
Moreover, 
\begin{align*}
E_{j, \delta}^{+} &= E_{j, \delta}(0) = E^* + \eta^- \delta + O(\delta^{2}), \\
E_{j+1, \delta}^{-} &= E_{j+1, \delta}(0)= E^* + \eta^+ \delta + O(\delta^{2}),
\end{align*} 
in which
\be \label{eq-eta}
\eta^-= \frac{a_2+ \sqrt{a_2^2 - a_1a_3}}{a_1}, \quad \eta^+= \frac{a_2- \sqrt{a_2^2 - a_1a_3}}{a_1}. 
\ee
\end{thm}

\noindent \textbf{Proof.} 
Note that 
$$
D(E, \delta) = 2+ a_1 (E-E^*)^2 + 2a_2 (E-E^*)\delta + a_3 \delta^2 + O(E-E^*)^3+ O(\delta^3).
$$
Solving $D(E, \delta) =2$ gives
$$
E= E^*+ \frac{a_2\pm \sqrt{a_2^2 - a_1a_3}}{a_1} \delta + O(\delta^{2}).
$$
Hence it is clear that $D(E, \delta) >2$ holds for $E \in (E^*+ \frac{a_2- \sqrt{a_2^2 - a_1a_3}}{2a_1} \delta, E^*+ \frac{a_2+\sqrt{a_2^2 - a_1a_3}}{2a_1} \delta )$.
The opening of the band gap follows by Lemma \ref{lem-12}.

Now for $\delta >0$, we have
\begin{align*}
E_{j, \delta}^{+} &= E_{j, \delta}(0)= E^*+ \frac{a_2+ \sqrt{a_2^2 - a_1a_3}}{a_1} \delta + O(\delta^{2}) = E^* + \eta^- \delta + O(\delta^{2})\\
E_{j+1, \delta}^{-} &= E_{j+1, \delta}(0)= E^*+ \frac{a_2- \sqrt{a_2^2 - a_1a_3}}{a_1} \delta + O(\delta^{2})=E^* + \eta^+ \delta + O(\delta^{2}).
\end{align*} 
This completes the proof.    \qed

\medskip

Before we end this section, we present scenarios for which the assumption (\ref{eq-assume1}) holds.
\begin{prop}
Let $\tilde \mu, \Tilde{\varepsilon}$ be such that $ \|\tilde \mu \|_{L^{\infty}} +\|\tilde \varepsilon \|_{L^{\infty}} =1 $. If $\tilde \mu \geq 0, \tilde \varepsilon \geq 0$, then there holds
$
a_2^2 - a_1 a_3 \geq 0.
$
\end{prop}

\noindent \textbf{Proof.}  Since $\tilde \mu \geq 0, \tilde \varepsilon \geq 0$, it follows that
$$
\int_{0}^1v^T(x)F(x)v(x) dx > 0, \quad \int_{0}^1u^T(x)F(x)u(x) dx > 0,
$$
and we can derive the following Cauchy-Schwartz type inequality
$$
\left|\int_{0}^1u^T(x)F(x)v(x) dx \right|^2 \leq \left|\int_{0}^1v^T(x)F(x)v(x) dx \cdot \int_{0}^1u^T(x)F(x)u(x) dx \right|. 
$$ 
Let
\begin{eqnarray*}
& |u|_1= \left(\int_{0}^1u^TWu dx\right)^{1/2}, \quad |v|_1 = \left(\int_{0}^1v^TWv dx\right)^{1/2}, \\
& |u|_2=\left(\int_{0}^1u^TFu dx \right)^{1/2}, \quad  |v|_2 = \left(\int_{0}^1v^TFv dx\right)^{1/2}, 
\end{eqnarray*}
and define
\begin{align*}
t_1 =& \frac{\int_{0}^1u^TWv dx}{(\int_{0}^1v^TWv dx)^{1/2}  \cdot (\int_{0}^1u^TWu dx)^{1/2}}=
 \frac{\int_{0}^1u^TWv dx}{|v|_1\cdot |u|_1} \\
t_2 =& \frac{\int_{0}^1u^TFv dx}{(\int_{0}^1v^TFv)^{1/2}  \cdot (\int_{0}^1u^TFu dx)^{1/2}}=
\frac{\int_{0}^1u^TFv dx}{|v|_2\cdot |u|_2}.
\end{align*} 
Then it is clear that $|t_1| \leq 1, |t_2| \leq 1$.
We obtain
\begin{align*}
\frac{1}{2}a_2= \frac{1}{2}\frac{\partial^2 D}{\partial E \partial \delta}(E^*, 0)  &=  t_1t_2\cdot |u|_1 \cdot |v|_1\cdot |u|_2 \cdot |v|_2- \frac{1}{2}|u|_2^2 \cdot |v|_1^2 -\frac{1}{2}|u|_1^2 \cdot |v|_2^2  \\
&=
 |u|_1 \cdot |v|_1\cdot |u|_2 \cdot |v|_2 \cdot \left( t_1t_2 - \frac{|u|_2 \cdot |v|_1}{2|u|_1\cdot |v|_2} -\frac{|u|_1\cdot |v|_2}{2|u|_2\cdot |v|_1}\right).
\end{align*} 
On the other hand
\begin{align*}
\frac{1}{2}a_1= \frac{1}{2}\frac{\partial^2 D}{\partial E^2}(E^*, 0)  &= ( t_1^2 -1)  \cdot |u|_1^2 \cdot|v|_1^2, \\
\frac{1}{2}a_3= \frac{1}{2} \frac{\partial^2 D}{\partial \delta^2}(E^*, 0) &= (t_2^2 -1) \cdot |u|_2^2 \cdot|v|_2^2.
\end{align*} 
Using the inequality $$(1-t_1t_2)^2 \geq (1-t_1^2)(1-t_2^2),$$ and 
$$
|t_1t_2 - \frac{|u|_2 \cdot |v|_1}{2|u|_1\cdot |v|_2} -\frac{|u|_1\cdot |v|_2}{2|u|_2\cdot |v|_1} | \geq  1-t_1t_2, 
$$
we can conclude that $a_2^2 - a_1 a_3 \geq 0$.   \qed

\subsection{Mode decomposition in the band gap} \label{sec-3-3}
In this subsection we assume that  (\ref{eq-assume1}) holds and therefore a band gap between the $j$-th and the $(j+1)$-th band is opened for the perturbed operator. We are interested in the modes in the band gap.  
According to Section \ref{sec-gapmode}, we construct the modes
\begin{eqnarray}
\phi_{E, 1, \delta} &=& \psi_{E, 2, \delta}(1) \psi_{E,2, \delta} + (\lambda_{E, 1, \delta} - \psi_{E, 1, \delta}(1)) \psi_{E,1, \delta}, \label{eq-phi_E1delta}  \\
\phi_{E, 2, \delta} &=& \psi_{E, 2, \delta}(1) \psi_{E,2, \delta} + (\lambda_{E, 2, \delta} - \psi_{E, 1, \delta}(1)) \psi_{E,1, \delta}. \label{eq-phi_E1delta} 
\end{eqnarray}
The former decays exponentially to $0$ as  $x \to +\infty$, and the latter decays exponentially to $0$ $x \to -\infty$. We have the following useful properties on the coefficients of the modes in the gap.

\begin{lem} \label{lem-egap}
let $E$ be in the band gap $(E_{j, \delta}^+, E_{j+1, \delta}^-)$, say
$\tau \eta^- \delta < {E-E^*} < \tau \eta^+\delta $   
for some $0< \tau <1$. Here $\eta^{\pm}$ are defined in (\ref{eq-eta}). 
Then we have
\begin{align*}
\lambda_{E, 1, \delta}& = 1-\sqrt{a_1(E-E^*)^2 + 2a_2 (E-E^*)\delta + a_3 \delta^2} +O(\delta^2), \\
\lambda_{E, 2, \delta}&
=1+ \sqrt{a_1(E-E^*)^2 + 2a_2 (E-E^*)\delta + a_3 \delta^2} +O(\delta^2). 
\end{align*}
and 
\begin{align*}
\psi_{E, 1, \delta}(1) &= 1+ \beta_1 (E- E^*)+ \tilde \beta_1 \delta + O(E-E^*)^2+ O(\delta^2);\\
\psi_{E, 2, \delta}(1) &= \beta_2 (E- E^*)+ \tilde \beta_2 \delta + O(E-E^*)^2+ O(\delta^2)
\end{align*}
where 
\begin{align*}
\beta_1 &= \int_{0}^1 v^T W udx, \quad \tilde \beta_1 = \int_{0}^1 v^T F udx, \\
\beta_2 &= \int_{0}^1 v^T W vdx, \quad \tilde \beta_2 = \int_{0}^1 v^T F vdx.
\end{align*}
\end{lem}

Proof. First note that $a_1 <0$ (by Lemma \ref{lem-a1}) and that
$$
D(E, \delta) = 2+ a_1 (E-E^*)^2 + 2a_2 (E-E^*)\delta + a_3 \delta^2 + O(E-E^*)^3+ O(\delta^3) .
$$
we have $$a_1 (E-E^*)^2 + 2a_2 (E-E^*)\delta + a_3 \delta^2 >0.$$
Then the two eigenvalues for the matrix $M(E, \delta)$ are given by
\begin{align*}
\lambda_{E, 1, \delta}& = \frac{D(E, \delta) - \sqrt{D(E, \delta)^2 -4}}{2}
=1-\sqrt{a_1(E-E^*)^2 + 2a_2 (E-E^*)\delta + a_3 \delta^2} +O(\delta^2), \\
\lambda_{E, 2, \delta}& = \frac{D(E, \delta) + \sqrt{D(E, \delta)^2 -4}}{2}
=1+ \sqrt{a_1(E-E^*)^2 + 2a_2 (E-E^*)\delta + a_3 \delta^2} +O(\delta^2). 
\end{align*}
The rest is a direct consequence of Taylor expansion and the formulas (\ref{eq-perturb1})-(\ref{eq-perturb2}) and (\ref{eq-Q}).

\subsection{Existence of an interface mode for the perturbed system}

For fixed $\delta >0$,  denote
\begin{align*}
\varepsilon_{\delta, \pm}(x)&= \varepsilon(x) \pm \delta \tilde \varepsilon(x) ,\quad 
\mu_{\delta, \pm}(x)= \mu(x)\pm \delta \tilde \mu(x),\\
\Lo_{\delta, \pm} \psi(x)&= -\frac{1}{\varepsilon_{\delta, \pm}(x)} \frac{d}{dx}\left(\frac{1}{\mu_{\delta,\pm}(x)} \frac{d\psi}{dx}\right). 
\end{align*}
We also define 
$$
\varepsilon_{\delta}(x)= 
\begin{cases}
\varepsilon(x) -\delta \tilde \varepsilon(x) , \quad x<0,\\
\varepsilon(x) +\delta \tilde \varepsilon(x), \quad x>0,
\end{cases}
\quad \mu_{\delta}(x)= 
\begin{cases}
\mu(x) -\delta \tilde \mu(x) , \quad x<0,\\
\mu(x)+\delta \tilde \mu(x), \quad x>0,
\end{cases}
$$
and the associated differential operator 
$$
\tilde \Lo_{\delta} \psi(x)= -\frac{1}{\varepsilon_{\delta}(x)}
\frac{d}{dx}\left(\frac{1}{\mu_{\delta}(x)} \frac{d\psi}{dx}\right), \quad \mbox{or equivalently}\,\, \tilde \Lo_{\delta}= 
\begin{cases}
\Lo_{\delta, -} , \quad x<0;\\
\Lo_{\delta, +}, \quad x>0.
\end{cases}
$$

Assume that the operator $\Lo_{0}$ attains a Dirac point $(k^*, E^*)$ in its band structure, which occurs at the intersection of the $j$-th and $(j+1)$-th band. 
Without loss of generality, we assume that  $k^*=0$. We shall make the following assumption on the perturbation (examples where the above assumption holds shall be given at the end of this section).
\begin{asump} \label{assump1}
The following assumption holds: 
either $\tilde \beta_1 =:\int_{0}^1u^TFv dx \geq 0$ 
and 
$$
\left(\int_{0}^1u^TFv dx\right)^2 > \left(\int_{0}^1v^TFv dx \right)\cdot \left(\int_{0}^1u^TFu dx \right),
$$
or 
$\tilde \beta_1 < 0$ and  
$$
\left(\int_{0}^1u^TFv dx\right)^2 >2 \left(\int_{0}^1v^TFv dx \right)\cdot \left(\int_{0}^1u^TFu dx \right),
$$
where $F$ is the matrix and the vectors $u$ and $v$ are given in (\ref{eq-F1}) and \eqref{eq-uv}, respectively.
\end{asump}
%
Recall that
$$
\frac{a_3}{2}=\frac{1}{2}\frac{\partial^2 D(E, \delta)}{\partial \delta^2}(E^*, 0)  =\left(\int_{0}^1u^TFv dx\right)^2 -
\left(\int_{0}^1v^TFv dx \right)\cdot \left(\int_{0}^1u^TFu dx \right).
$$
The above assumption implies that $a_3 >0$. Recall that $a_1 <0$ (by Lemma \ref{lem-a1}). We see that 
the inequality (\ref{eq-assume1}) holds.
In light of Theorem \ref{thm-bandgap},  the perturbation $\tilde \varepsilon, \tilde \mu$ will create a band gap $(E_{j, \delta, \pm}^{+}, E_{j+1, \delta, \pm}^{-})$
at $E^*$ for the operators $\Lo_{\delta, \pm}$.
In addition,
\begin{align*}
E_{j, \delta, \pm}^{+} &= E^* \pm \eta^- \delta + O(\delta^{3/2}), \\
E_{j+1, \delta, \pm}^{-} &= E^* \pm \eta^+ \delta + O(\delta^{3/2}),
\end{align*} 
where $\eta^{\pm}$ are defined in (\ref{eq-eta}). 

\begin{lem}
Under Assumption \ref{assump1}, the intersection of two band gaps $(E_{j, \delta, +}^{+}, E_{j+1, \delta, +}^{-}) \cap (E_{j, \delta, -}^{+}, E_{j+1, \delta, -}^{-})$ for the two operators $\Lo_{\delta, +}$ and $\Lo_{\delta, -}$ is not empty.
\end{lem}
\noindent\textbf{Proof. }
Since $a_1 <0, a_3 >0$, we have $\eta^- <0 < \eta^+$. 
If $a_2 <0$, it follows that $ |\eta^-| > |\eta^+|$, and consequently  
$$
(E_{j, \delta, +}^{+}, E_{j+1, \delta, +}^{-}) \cap (E_{j, \delta, -}^{+}, E_{j+1, \delta, -}^{-}) = (E^*- \eta^+ \delta + O(\delta^{3/2}), E^*+ \eta^+ \delta + O(\delta^{3/2})). 
$$
On the other hand, if $a_2 >0$, then $|\eta^-| < |\eta^+|$
and there holds
$$
(E_{j, \delta, +}^{+}, E_{j+1, \delta, +}^{-}) \cap (E_{j, \delta, -}^{+}, E_{j+1, \delta, -}^{-}) = (E^*+ \eta^- \delta + O(\delta^{3/2}), E^*- \eta^- \delta + O(\delta^{3/2})). 
$$
\qed

Next we investigate the existence of an interface mode in the band gap for the operator $ \tilde \Lo_{\delta}$. 

\begin{thm}
Assume that the operator $\Lo_{0}$ attains a Dirac point $(k^*=0, E^*)$ in its band structure, which occurs at the intersection of the $j$-th and $(j+1)$-th band. Further assume that Assumption \ref{assump1} holds for the perturbation, then there exists an interface mode for the operator $ \tilde \Lo_{\delta}$ for $\delta$ sufficiently small. The same conclusion also holds when the Dirac point occurs at $k^*=\pi$. 
\end{thm}

\noindent\textbf{Proof.} Without loss of generality, we only consider the case $k^*=0$. We may further restrict to the case when $a_2 <0$ since the case $a_2 >0$ can be treated similarly.  First we have
$$
(E_{j, \delta, +}^{+}, E_{j+1, \delta, +}^{-}) \cap (E_{j, \delta, -}^{+}, E_{j+1, \delta, -}^{-}) = \big(E^*- \eta^+ \delta + O(\delta^{3/2}), E^*+ \eta^+ \delta + O(\delta^{3/2})\big). 
$$
For the operator $\Lo_{\delta, +}$, we construct mode of the type \eqref{eq-phi_E1delta} that decays exponentially as  $x \to +\infty$: 
$$
\phi_{E, 1, \delta} = \psi_{E, 2, \delta}(1) \psi_{E,2, \delta} + (\lambda_{E, 1, \delta} - \psi_{E, 1, \delta}(1)) \psi_{E,1, \delta}.  
$$
Define
\[
\xi_{R, \delta}(E) =\frac{\psi_{E, 2, \delta}(1)}{\lambda_{E, 1, \delta} - \psi_{E, 1, \delta}(1)}. 
\]
By Lemma \ref{lem-egap}, we have
\begin{eqnarray*}
\xi_{R, \delta}(E)
= \frac{ \beta_2(E-E^*) + \tilde \beta_2 \delta +O(\delta^2)}{-\sqrt{a_1 (E-E^*)^2 +2a_2 (E-E^*) \delta + a_3 \delta^2} -  \beta_1(E-E^*) -\tilde \beta_1 \delta +O(\delta^2)}.
\end{eqnarray*}
Recall that 
$$
\tilde \beta_1 =\int_{0}^1u^TFv dx, \quad \frac{a_3}{2} =\left(\int_{0}^1u^TFv dx\right)^2 -\left(\int_{0}^1v^TFv dx \right)\cdot \left(\int_{0}^1u^TFu dx \right). 
$$
Under Assumption \ref{assump1}, we can find $0< \tau_1 <\eta^+$ such that for all $E$ satisfying $|E-E^*|\leq \tau_1 \delta$,
$$
|\sqrt{a_1 (E-E^*)^2 +2a_2 (E-E^*) \delta + a_3 \delta^2} +  \beta_1(E-E^*) +\tilde \beta_1 \delta| \geq c_1 \delta 
$$
for some constant $c_1>0$. Therefore $\xi_{R, \delta}(E)$ is well-defined for 
$E$ satisfying $|E-E^*|\leq \tau_1 \delta$ and for $\delta$ sufficiently small.

For the operator $\Lo_{\delta, -}$,  we let
$$
\phi_{E, 2, -\delta} = \psi_{E, 2, -\delta}(1) \psi_{E,2, -\delta} + (\lambda_{E, 2, -\delta} - \psi_{E, 1, -\delta}(1)) \psi_{E,1, -\delta},
$$
and define
\begin{eqnarray*}
\xi_{L, -\delta}(E) = \frac{\psi_{E, 2, -\delta}(1)}{\lambda_{E, 2, -\delta} - \psi_{E, 1, -\delta}(1)}.
\end{eqnarray*}
Using Lemma \ref{lem-egap} again, we have
\[
\xi_{L, -\delta}(E) =  \frac{ \beta_2(E-E^*) - \tilde \beta_2 \delta +O(\delta^2)}{\sqrt{a_1 (E-E^*)^2 -2a_2 (E-E^*) \delta + a_3 \delta^2} -  \beta_1(E-E^*) +\tilde \beta_1 \delta +O(\delta^2)}.
\]
Similar to the previous argument, $\xi_{L, -\delta}(E)$ is well-defined for 
$E$ satisfying $|E-E^*|\leq \tau_1 \delta$ and for $\delta$ sufficiently small.
%

Now if $u$ is an interface modes for $\Lo_{\delta}$, by Lemma \ref{lem-gapmode} there holds
$$
\begin{pmatrix}
u(0)\\
\frac{1}{\mu(0)}u'(0)
\end{pmatrix} = c_1 \begin{pmatrix}
\psi_{E, 2, \delta}(1)\\
\lambda_{E, 1, \delta} - \psi_{E, 1, \delta}(1)
\end{pmatrix}=
 c_2 \begin{pmatrix}
\psi_{E, 2, -\delta}(1)\\
\lambda_{E, 2, -\delta} - \psi_{E, 1, -\delta}(1)
\end{pmatrix}.
$$
Therefore, there exists an interface mode at energy level $E$ if and only if 
$$
\xi_{R, \delta}(E) = \xi_{L, -\delta}(E). 
$$ 
Let $t= E-E^*$, then $-\eta^+ \delta < t < \eta^+ \delta.$
We define the function
\begin{align*}
g(t) =\xi_{R, \delta}(E) - \xi_{L, -\delta}(E) = &\frac{ \beta_2t + \tilde \beta_2 \delta +O(\delta^2)}{-\sqrt{a_1 t^2 +2a_2 t \delta + a_3 \delta^2} -  \beta_1t -\tilde \beta_1 \delta +O(\delta^2)}  \\
& - \frac{ \beta_2t - \tilde \beta_2 \delta +O(\delta^2)}{\sqrt{a_1 t^2 -2a_2 t \delta + a_3 \delta^2} - \beta_1t +\tilde \beta_1 \delta +O(\delta^2)}.
\end{align*}
It is clear that 
$$
g(t) = \tilde g(t) +O(\delta)
$$
where 
$$
\tilde g(t) = \frac{ \beta_2t + \tilde \beta_2 \delta }{-\sqrt{a_1 t^2 +2a_2 t \delta + a_3 \delta^2} -  \beta_1t -\tilde \beta_1 \delta }  
 + \frac{ -\beta_2t + \tilde \beta_2 \delta }{\sqrt{a_1 t^2 -2a_2 t \delta + a_3 \delta^2} -  \beta_1t +\tilde \beta_1 \delta}.
$$
One can check directly that $\tilde g(t)$ is odd, i.e. $\tilde g(t) + \tilde g(-t) =0 $.

We now choose $0<\tau_2<1$ such that $|\tau_2 \eta^+|<\tau_1$ and 
$\tilde g(\tau_2 \eta^+ \delta) \neq 0$. It then follows that
$$
\tilde g(\tau_2 \eta^+ \delta) \cdot \tilde g(-\tau_2 \eta^+ \delta) <0 
$$ 
Note that the value of $\tilde g(\tau_2 \eta^+ \delta)$ is independent of $\delta$. 
For $\delta$ small enough, we have
$$ 
g(\tau_2 \eta^+ \delta) \cdot g(-\tau_2 \eta^+ \delta) <0. 
$$
Hence there exists a root to $g(t)=0$ in the interval $(-\tau_2 \eta^+ \delta, \tau_2 \eta^+ \delta )$. By our previous argument, this root gives the existence of an interface mode with exactly the same energy level. \qed

\begin{rem}
Bulk-interface correspondence is not formulated in the above theorem for the existence of the interface mode. 
Also the stability of the interface mode under perturbations that are not small is a subtle issue and is not discussed here. One possible formulation of the stability is to show the persistence of the interface mode under a continuous family of  perturbations to the operator $\tilde \Lo_{\delta}$ such that band-gap structure of the bands below the gap for the two periodic operators on the two semi-infinite intervals $x<0$ and $x>0$ is kept unchanged in the process. We leave this as a future work. 
\end{rem}

Finally, we investigate the scenario where Assumption \ref{assump1} holds.

\begin{lem}
Assume that $\mu$ and $\varepsilon$ are even functions. Further assume that 
$\tilde{\mu}$ and $\tilde \varepsilon$ are odd, then
$$
\left(\int_{0}^1v^TFv dx \right)\cdot \left(\int_{0}^1u^TFu dx \right)=0, \quad  \frac{a_3}{2} =\left(\int_{0}^1u^TFv dx\right)^2.  
$$
Moreover, we can choose $\tilde{\mu}$ and $\tilde \varepsilon$ such that
$$
\int_{0}^1u^TFv dx \neq 0. 
$$
\end{lem}
\noindent\textbf{Proof.}  For clarity of presentation, we set  
$\tilde{\mu}=0$. Recall that 
$$
u(x) = \psi_{E^*,1, 0}(x), \quad v(x) = \psi_{E^*,2, 0}(x).
$$
Thus $u$ is an even function and $v$ an odd function.  
Since $\Psi_{E^*, 0}=Id$, we have $\Psi_{E^*, 0}(x+1) = \Psi_{E^*, 0}(x)$. It follows that
$$
u(x+1) =u(x), \quad v(x+1) =v(x).
$$
Therefore,
$$
\int_{0}^1u^TFu dx = \int_{0}^1 \psi_{E^*,1}^2(x) E \tilde \varepsilon(x) dx =\int_{-1/2}^{1/2} \psi_{E^*,1}^2(x) E \tilde \varepsilon(x) dx =0,
$$
and that 
\begin{align*}
\int_{0}^1u^TFv dx &= 
\int_{0}^1 \psi_{E^*,1}(x)\psi_{E^*,2}(x) E \tilde \varepsilon(x) dx =\int_{-1/2}^{1/2} \psi_{E^*,1}(x)\psi_{E^*,2}(x) E \tilde \varepsilon(x) dx\\
&=2E\int_{0}^{1/2} \psi_{E^*,1}(x)\psi_{E^*,2}(x) \tilde \varepsilon(x) dx.
\end{align*}
It is clear that we can choose $\tilde \varepsilon(x)$ to make $\int_{0}^1u^TFv dx  \neq 0$. This completes the proof of the lemma.  \qed

\section{Photonic structures with inversion symmetry}\label{sec:inv_sym}
In this section, we assume that the time-reversal symmetric photonic structure (\ref{eq-photonic}) attains additional inversion symmetry with $\varepsilon(x)= \varepsilon(1-x)$, $\mu(x) = \mu(1-x)$,
or equivalently, $\varepsilon(x)= \varepsilon(-x)$, $\mu(x) = \mu(-x)$.
Such topological structures were investigated in \cite{chan-14} and it was shown that localized mode exists at the interface of the two semi-infinite periodic structures with different bulk topological indices.
Inspired by this work, we would like to provide a rigorous theory for the existence of an interface mode for such a  structure
and its connection 
to the bulk topological index, which is defined via the quantized Zak phase.
In addition, we investigate the stability of the interface mode under perturbations that are not necessarily small.

\subsection{Bloch modes and parity}
\begin{lem}\label{lem-psi_sym}
Under inversion symmetry, if $\varphi_{j,k}(x)$ is a Bloch mode for the $j$-th band with Bloch wavenumber $k$, then $\varphi_{j,k}(-x)$ is a Bloch mode for the Bloch wavenumber $-k$. 
\end{lem}
\noindent\textbf{Proof.} Let $E=E_j(k)$. The Bloch mode $\varphi_{j,k}(x)$ solves
$$
\begin{cases}
(\Lo -E) \varphi_{j,k} =0, \\
\varphi_{j,k}(x+1 ) = e^{ik}\varphi_{j,k}(x).
\end{cases}
$$
Let $v(x)=\varphi_{j,k}(-x) $. We have
$$
\begin{cases}
(\Lo -E) v =0, \\
v(x+1 ) = \varphi_{j,k}(-x-1) = \varphi_{j,k}(-x)e^{-ik} = v(x)e^{-ik},
\end{cases}
$$
which shows that $v(x)=\varphi_{j,k}(-x) $ is a Bloch mode for the Bloch vector $-k$.  \qed \\

\begin{lem} \label{lem-42}
Under inversion symmetry, the Bloch modes $\varphi_{j,k}$ are even or odd when $k=0$ or $\pi$ over an isolated band
 $E_j(k)$. In addition, for $k=0$ or $\pi$, there holds $\varphi_{j,k}=c \,\psi_{E, 1}$  or $\varphi_{j,k} = c \, \psi_{E, 2}$ for certain constant $c$ depending on whether $\varphi_{j,k}$ is even
or odd.
\end{lem}
\noindent\textbf{Proof.} Consider the Bloch mode $\phi_{j,0}$ for $k=0$ which solves the following equations
$$
(\Lo -E_j(0))\phi_{j,0} =0, \quad \phi_{j,0}(x+1) =  \phi_{j,0}(x).
$$
Let $v(x) = \phi_{j,0}(-x)$, then there holds
$$
(\Lo -E_j(0))v =0, \quad v(x+1) =  v(x).
$$
Thus $v(x) = \phi_{j,0}(-x)$ is also a Bloch mode for $k=0$. 
Since the multiplicity of the Bloch mode for $k=0$ is one (Propositon \ref{lem-multi}) and that $\phi_{j,0}$ is real-valued, it follows that $v(x)= \pm \phi_{j,0}(x)$, i.e.
$$
\phi_{j,0}(-x) = \pm \phi_{j,0}(x).
$$
Note that if $\phi_{j,0}$ is even, then $\phi_{j,0}'(0) =0$ and $\phi_{j,0}= c \psi_{E, 1}$
for some constant $c$. Similarly, if $\phi_{j,0}$ is odd, then 
$\phi_{j,0}= c \psi_{E, 2}$
for some constant $c$. A parallel argument leads to the conclusion for the Bloch mode $\phi_{j,\pi}$.  \qed

\begin{defn}
We call that the Bloch mode $\varphi_{j,k}$ attains an even-parity (odd-parity) if $\varphi_{j,k}$ is an even (odd) function.
\end{defn}


Next we investigate the change of parity for the Bloch modes at $k=0$ or $\pi$ when the energy crosses a band gap. A crucial tool we used is the oscillation theory 
for  Sturm-Liouville operators, see for instance \cite{weidmann-1987}. To be more precise, 
let us denote $E_j^P, E_j^S, E_j^D, E_j^N$ the $j$-th eigenvalues of the operator $\Lo$ restricted to the unit cell $[0, 1]$ with the following boundary conditions respectively: 
\begin{enumerate}
\item [(i)]
Periodic boundary conditions: $u(1)=u(0)$, \,\,\, $u'(1)=u'(0)$;
\item [(ii)]
Semi-periodic boundary conditions: $u(1)=-u(0)$, \,\,\, $u'(1)=-u'(0)$;
\item [(iii)]
Dirchilet boundary conditions: $u(1)=u(0)=0$; 
\item [(iv)]
Nuemann boundary conditions: $u'(1)=u'(0)=0$
\end{enumerate}
We have the following theorem on the eigenvalues above, see for instance Theorem 13.10 in \cite{weidmann-1987}.
\begin{thm}\label{thm-interlace}
The eigenvalues $E_j^P, E_j^S, E_j^D, E_j^N$ $(j = 1, 2, 3, \cdots )$ attain the following interlacing property:
\begin{align*}
E_1^N & \leq E_1^P < E_1^S \leq  \{E_2^N, E_1^D \} \leq E_2^S < E_2^P \leq  \{E_3^N, E_2^D \} \leq \cdots  \\
& \leq E_{2n-1}^P < E_{2n-1}^S \leq  \{E_{2n}^N, E_{2n-1}^D \} \leq E_{2n}^S < E_{2n}^P  \leq  \{E_{2n+1}^N, E_{2n}^D \}  \leq E_{2n+1}^P < \cdots
\end{align*}
\end{thm}

Based on the above theorem, we are able to show the change of parity for the Bloch modes across the band gap, which is stated in the theorem below.
\begin{thm} \label{thm-parity_change}
Assume that the $j$-th band is isolated, then the Bloch modes at $(k, E_j^+)$ and $(k,E_{j+1}^-)$ attain different parity, where $k=0$ or $\pi$.
\end{thm}
\noindent\textbf{Proof.} Without loss of generality, we assume that $k=0$ so that $E_j^+ =E_j(0)$,   $E_{j+1}^-=  E_{j+1}(0)$, and 
the Bloch mode $\varphi_{j ,0}$ at $(0, E_j^+)$ is even. Then $\varphi_{j ,0}$ satisfies the following boundary value problem:
$$
\begin{cases}
(\Lo -E_j^+)\varphi_{j ,0} = 0, \\
\varphi_{j ,0}(0) = \varphi_{j ,0}(1), \\
\varphi_{j ,0}'(0) = \varphi_{j ,0}'(1)=0.
\end{cases}
$$
Hence $E_j^+$ is a common eigenvalue to the operator $\Lo$ for both the periodic boundary condition and the Neumann boundary condition. We prove by contradiction that $\varphi_{j+1 ,0}$ is odd. Otherwise, if $\varphi_{j+1 ,0}$ is even, then $E_{j+1}^-$ is also a common eigenvalue to the operator $\Lo$ 
for both the periodic boundary condition and the Neumann boundary condition.
Note that $E_j^+ < E_{j+1}^-$ are two neighboring eigenvalues to $\Lo$ with the periodic boundary condition. We either have $E_j^+ =E_{2n-1}^P, E_{j+1}^-=E_{2n}^P $ or $E_j^+ =E_{2n}^P, E_{j+1}^-=E_{2n+1}^P$ for some integer $n$. By Theorem \ref{thm-interlace}, the former is impossible since there is no eigenvalue to $\Lo$ with the semi-periodic boundary condition inside the band gap.
The latter is also impossible since both $E_j^+ , E_{j+1}^-$ are eigenvalues to $\Lo$ with the Neumann boundary condition. This contradiction proves that $\varphi_{j+1 ,0}$ should be an odd-parity mode and this completes the proof for the case $k=0$. The case $k=\pi$ can be proved in a similar manner. \qed

\subsection{Zak phase}
\subsubsection{Zak phase for an isolated band}\label{sec-Zak1_inv_sym} 
Following Section \ref{sec-bloch_states}, we construct the Bloch modes for an isolated band $E_j(k)$  as follows:
\be  \label{eq-psi-inversion}
\varphi_{j,k}(x) = 
\begin{cases}
\frac{ \phi_{j,k}(x)}{\|\phi_{j,k}\|_{X}},  0\leq  k \leq  \pi, \quad \phi_{j,k} \not\equiv 0, \\
\frac{i \psi_{E, 2}(x)}{\|\psi_{E, 2}\|_{X}},  \,\, k\in \{0, \pi\} \,\,\,\mbox{and}\,\,\,\phi_{j,k} \equiv 0,\\
 \varphi_{j,-k}(-x), \quad -\pi < k <0.
\end{cases}
\ee
The periodic part of $\varphi_{j,k}$ is given by $u_{j,k}(x) = \varphi_{j,k}(x) e^{-ikx}.$
From Lemma \ref{lem-psi_sym} we have 
$$
u_{j, k}(x) = u_{j, -k}(-x), \quad -\pi<k <0.
$$
We calculate the Zak phase using the formula \eqref{eq-zak1}.
First, note that
\begin{align*}
\int_{-\pi}^{0} \left(\frac{\partial u_{j,k}}{\partial k}, u_{j, k}\right)_{X} \, dk
&= 
-\int_{0}^{\pi} \left(\frac{\partial u_{j,-k}}{\partial k}, u_{j, -k}\right)_{X} dk \\
&=
-\int_{0}^{\pi} \int_{0}^1 \frac{\partial u_{j,-k}(x)}{\partial k}\bar u_{j, -k}(x) \varepsilon(x) dx dk\\
&= -\int_{0}^{\pi} \int_{0}^1  \frac{\partial u_{j,k}(-x)}{\partial k}\bar u_{j, k}(-x) \varepsilon(-x) dx dk \\
& =
-\int_{0}^{\pi} \int_{0}^1  \frac{\partial u_{j,k}(1-x)}{\partial k}\bar u_{j, k}(1-x) \varepsilon(1-x) dx dk \\
&=
-\int_{0}^{\pi} \int_{0}^1 \frac{\partial u_{j,k}(x)}{\partial k}\bar u_{j, k}(x) \varepsilon(x) dx dk\\
&= 
-\int_{0}^{\pi}\left(\frac{\partial u_{j,k}}{\partial k}, u_{j, k}\right)_{X} \, dk.
\end{align*}
On the other hand, note that
\begin{align*} 
u_{j, 0^-}(x) &= \lim_{k \to 0^-} u_{j, k}(x) = \lim_{k \to 0^+} u_{j, -k}(x)= \lim_{k \to 0^+} u_{j, k}(-x) = u_{j, 0}(-x); \\
u_{j, (-\pi)^+}(x) &= \lim_{k \to (-\pi)^+} u_{j, k}(x) = \lim_{k \to \pi^-} u_{j, -k}(x)= \lim_{k \to \pi^-} u_{j, k}(-x) = u_{j, \pi}(-x).
\end{align*}
It follows that
\begin{align*} 
(u_{j, 0}, u_{j, 0^-})_X &= 
\int_{0}^{1/2} u_{j, 0}(x) \bar u_{j, 0}(-x)  \varepsilon(x) dx + \int_{1/2}^{1} u_{j, 0}(x) \bar u_{j, 0}(-x)  \varepsilon(x) dx\\
& = \int_{0}^{1/2} u_{j, 0}(x) \bar u_{j, 0}(-x)  \varepsilon(x) dx
+ \int_{-1/2}^{0} u_{j, 0}(x+1) \bar u_{j, 0}(-x-1)  \varepsilon(x+1) dx\\ 
& = \int_{0}^{1/2} u_{j, 0}(x) \bar u_{j, 0}(-x)  \varepsilon(x) dx
+ \int_{-1/2}^{0} u_{j, 0}(x) \bar u_{j, 0}(-x)  \varepsilon(x) dx \\
& = \int_{-1/2}^{1/2} u_{j, 0}(x) \bar u_{j, 0}(-x)  \varepsilon(x) dx 
 = \int_{-1/2}^{1/2} \varphi_{j,0}(x) \bar \varphi_{j,0}(-x)  \varepsilon(x) dx \\
& =\begin{cases}
1, \quad \mbox{if} \,\,\, \varphi_{j,0}(x) = \varphi_{j,0}(-x), \\ 
-1, \quad \mbox{if} \,\,\, \varphi_{j,0}(x) = -\varphi_{j,0}(-x).
\end{cases}
\end{align*}
Similarly, we have
\begin{equation}\label{eq-u_pi_inner_prod}
(  e^{-i2\pi x} u_{j, (-\pi)^+},u_{j, \pi})_X =\begin{cases}
1, \quad \mbox{if} \,\,\, \varphi_{j,\pi}(x) = \varphi_{j,\pi}(-x), \\ 
-1, \quad \mbox{if} \,\,\, \varphi_{j,\pi}(x) = -\varphi_{j,\pi}(-x).
\end{cases}
\end{equation}
Therefore, by substituting the above into the formula \eqref{eq-zak1}, the Zak phase for an isolated band can be characterized in the following theorem.
\begin{thm} \label{thm-zak1}
For a time-reversal symmetric periodic structure with inversion symmetry, the Zak phase for an isolated band $E_j(k)$ is given by
$$
\theta_j = \begin{cases}
0, \quad \mbox{if $\varphi_{j,0}(x) \; \mbox{and} \; \varphi_{j,\pi}(x)$ attain the same parity}, \\ 
\pi, \quad \mbox{if $\varphi_{j,0}(x) \;  \mbox{and} \; \varphi_{j,\pi}(x)$ attain different parity}.
\end{cases}
$$
\end{thm}

\subsubsection{Zak phase at the presence of Dirac point}\label{sec-Zak2_inv_sym}
Without loss of generality, we assume that the band $E_j(k)$ and $E_{j+1}(k)$ crosses at the Dirac point $(0, E_j^+)$.
%
From the discussions in Section \ref{sec-Zak1_inv_sym}, the Zak phase for the $j$-th band is
\begin{align*}
\theta_j &= - \Im \ln (u_{j, 0}, u_{j, 0-})_X - \Im \ln (  e^{-i2\pi x} u_{j, (-\pi)^+},u_{j, \pi})_X  \,\, mod\,\, 2\pi \\
& = - \Im \ln (\varphi_{j,0},\varphi_{j,0^-})_X - \Im \ln (\varphi_{j,(-\pi)^+},\varphi_{j,\pi})_X  \,\, mod\,\, 2\pi \\
& = - \Im  \ln (\varphi_{j,0}(x),\varphi_{j,0}(-x))_X - \Im  \ln (\varphi_{j,\pi}(-x), \varphi_{j,\pi}(x))_X  \,\, mod\,\, 2\pi.
\end{align*}
Similarly, the Zak phase for $j+1$-th band is
$$
\theta_{j+1}= - \Im  \ln (\varphi_{j+1,0}(x),\varphi_{j+1,0}(-x))_X - \Im  \ln (\varphi_{j+1,\pi}(-x),\varphi_{j+1,\pi})_X  \,\, mod\,\, 2\pi.
$$
In light of the relations $\varphi_{j+1, 0}(x) = -\overline{\varphi_{j,0}(x)}$ and $\varphi_{j+1,0}(-x) = -\overline{\varphi_{j,0}(-x)}$ given in Section \ref{sec-Zak2}, it follows that
$$
\theta_j + \theta_{j+1} = - \Im \ln (\varphi_{j,(-\pi)^+},\varphi_{j,\pi})_X - \Im \ln (\varphi_{j+1,(-\pi)^+},\varphi_{j+1,\pi})_X  \,\, mod\,\, 2\pi.
$$
Now the following theorem follows by using \eqref{eq-u_pi_inner_prod}.
\begin{thm} \label{thm-zak-dirac2}
For a periodic structure with inversion symmetry, if the band $E_j(k)$ and $E_{j+1}(k)$ cross at the Dirac point $(k=0, E_j^+)$ and they do not cross with other bands in the spectrum,
then 
\be
\theta_j + \theta_{j+1} = \begin{cases}
0, \quad \mbox{if $\varphi_{j,\pi}(x), \varphi_{j+1,\pi}(x)$ attain the same parity}; \\
\pi, \quad \mbox{if $\varphi_{j,\pi}(x), \varphi_{j+1,\pi}(x)$ attain different parities}.
\end{cases}
\ee
If the $j$-th band and the $j+1$-th band cross at the Dirac point $(k=\pi, E_j^+)$, then
\be
\theta_j + \theta_{j+1} = \begin{cases}
0, \quad \mbox{if $\varphi_{j,0}(x), \varphi_{j+1,0}(x)$ attain the same parity}; \\
\pi, \quad \mbox{if $\varphi_{j,0}(x), \varphi_{j+1,0}(x)$ attain different parities}.
\end{cases}
\ee
\end{thm}

\subsection{Interface modes and bulk topological indices} 

\subsubsection{Bulk topological indices}
Assume that the band $E_j(k)$ does not cross with $E_{j+1}(k)$ so that there is a gap between the two bands.
We define an index for the band $E_j(k)$ of the periodic structure as
\begin{equation}\label{eq-zeta}
\gamma_j=(-1)^{j+\ell-1} e^{i \sum_{m=1}^{j} \theta_m},
\end{equation}
in which $\theta_m$ is  the Zak phase for the band $E_m(k)$, and  $\ell$ is the number of Dirac points below the band $E_j(k)$.
The relation between the parity of the Bloch mode $\psi_{k,j}$ at band edge $(k, E_j^+)$ and the bulk index $\gamma_j$ is given in the following theorem.

\begin{thm}
The bulk topological index $\gamma_j$ only takes the values $\pm 1$. In addition,  for a given band $E_j(k)$,
$\gamma_j$  is  $1$ and $-1$  when the Bloch mode $\psi_{k,j}$ at band edge $(k,E_j^+)$ is even and odd  respectively.
\end{thm}

\noindent\textbf{Proof.}  Recall that for the first band, we have $E_1^-=E_1(0)$ and the associated Bloch mode is a constant function.
Since $E_1^+=E_1(\pi)$,
by virtue of Theorem \ref{thm-zak1}, $\gamma_1=1$ and $-1$ when the Bloch mode $\psi_{\pi,1}$  at $(\pi, E_1^+)$ is even and odd respectively.
Now we prove by induction and assume that the statement holds for the band $E_{n}(k)$ with $n<j$. If $E_{j-1}(k)$ does not cross with $E_{j}(k)$, 
then an application of Theorems \ref{thm-parity_change} and \ref{thm-zak1} yields $\gamma_j = - e^{i \theta_j} \gamma_{j-1} $,
where $\theta_j$ is $0$ or $\pi$.
Otherwise, if $E_{j-1}(k)$ and $E_{j}(k)$ cross at the Dirac point $(k, E_j^-)$, applying Theorem  \ref{thm-zak-dirac2} gives $\gamma_j = - e^{i (\theta_{j-1}+\theta_j)} \gamma_{j-2} $,
where $\theta_{j-1}+\theta_j=0$ or $\pi$. The proof is complete. \qed


\subsubsection{Mode decomposition in the band gap}
In this section, we consider the mode decomposition in the band gap under the inversion symmetry assumption for the underlying system. 
To be more specific, we assume that the $j$-th band gap is open and $E \in (E_j^+, E_{j+1}^-)$. 

\begin{prop}\label{prop-gap_mode_inv_sym}
If the periodic structure attains inversion symmetry and $u\in L^1_{loc}(\mathbf{R})$ is a solution to $\Lo u = Eu$,
then the following holds
for $E \in (E_j^+, E_{j+1}^-)$:
\begin{enumerate}
\item [(i)]
$\psi_{E, 2}(1) \neq 0, \quad \lambda_{E, 1} - \psi_{E, 1}(1) \neq 0, \quad \lambda_{E, 2} - \psi_{E, 1}(1) \neq 0$.
\item [(ii)]
$u(x) \to 0$ as $x \to +\infty$ and $|u(x)| \to \infty$ as $x \to -\infty$ 
if and only if $U(0)=c V_{E,1}$
for some constant $c$. Similarly, $u(x) \to 0$ as $x \to -\infty$ and $|u(x)| \to \infty$ as $x \to \infty$
if and only if $U(0)=c V_{E,2}$
for some constant $c$.
Here $U(0):=(u(0), \frac{1}{\mu(x)}u'(0))^T$ is the solution vector at $x=0$, $V_{E,1}$ and $V_{E,2}$ are eigenvectors defined by \eqref{eq-eigenvec}.
\end{enumerate}
\end{prop}

\noindent\textbf{Proof.} We first show that $\psi_{E, 2}(1) \neq 0$. Assume otherwise that $\psi_{E, 2}(1) = 0$.  Define $\psi(x) = \psi_{E, 2}(1-x)$. Then both
$\psi$ and $\psi_{E, 2}$ are solutions to the following boundary value problem:
$$
(\Lo -E)u=0, \quad u(0) = u(1)=0.
$$
Thus $\psi$ and $\psi_{E, 2}$ must be linearly dependent. Since both are real-valued, we see that $\psi = \pm \psi_{E, 2}$. It follows that $\psi'(1)= -\psi_{E, 2}'(0) = \pm \mu(0) $, and hence
$$
M(E)= 
\begin{pmatrix}
\psi_{E, 1}(1) &  \psi_{E, 2}(1) \\
\frac{1}{\mu(0)}\psi_{E, 1}'(1) &  \frac{1}{\mu(0)} \psi_{E, 2}'(1)
\end{pmatrix} = 
\begin{pmatrix}
\psi_{E, 1}(1) &  0 \\
\frac{1}{\mu(0)}\psi_{E, 1}'(1) &  \pm 1
\end{pmatrix}
$$
Therefore, we see that $\pm 1 \in \{\lambda_{E, 1}, \lambda_{E, 2}\}$, which is a contradiction
 to the fact that  $|\lambda_{E, 1}| <1$ and $|\lambda_{E, 2}| >1$ in the band gap. 

We next prove that $\lambda_{E, 1} - \psi_{E, 1}(1) \neq 0$. 
If $\lambda_{E, 1} - \psi_{E, 1}(1) = 0$, then by using $det (M(E)- \lambda_{E, 1})=0$, we have
$$
\psi_{E, 2}(1) \cdot \frac{1}{\mu(0)}\psi_{E, 1}'(1) =0,
$$
which yields $\psi_{E, 1}'(1) =0$.
Let $\psi(x) = \psi_{E, 1}(1-x)$, then both
$\psi$ and $\psi_{E, 1}$ are solutions to the following boundary value problems:
$$
(\Lo -E)u=0, \quad u'(0) = u'(1)=0. 
$$
Using the similar argument as above, we obtain $\psi = \pm \psi_{E, 1}$. This implies that
$$
\psi_{E, 1}(0) = \pm \psi_{E, 1}(1) = \pm 1. 
$$
Hence
$$
M(E)= 
\begin{pmatrix}
\psi_{E, 1}(1) &  \psi_{E, 2}(1) \\
\frac{1}{\mu(0)}\psi_{E, 1}'(1) &  \frac{1}{\mu(0)} \psi_{E, 2}'(1)
\end{pmatrix} = 
\begin{pmatrix}
\pm 1 &  \psi_{E, 2}(1) \\
0 &  \pm \frac{1}{\mu(0)} \psi_{E, 2}'(1)
\end{pmatrix}.
$$
Again, this leads to $\pm 1 \in \{\lambda_{E, 1}, \lambda_{E, 2}\}$, which contradicts to the fact that the eigenvalues are in the band gap.
This complete the proof of (i).
(ii) follows from Lemma \ref{lem-gapmode}. \qed

\medskip

%

We define two impedance functions $\xi_{R}(E)$ and $\xi_{L}(E)$ by letting
\begin{align}
\xi_{R}(E) = \frac{\psi_{E, 2}(1)}{\lambda_{E, 1} - \psi_{E, 1}(1)}, \quad
\xi_{L}(E) = \frac{\psi_{E, 2}(1)}{\lambda_{E, 2} - \psi_{E, 1}(1)}.
\end{align}

\begin{lem} \label{lem-edgemode2}
For a time-reversal symmetric periodic structure that attains inversion symmetry,
the following holds for $E \in (E_j^+, E_{j+1}^-)$:
\begin{enumerate}
\item [(i)]
If the Bloch mode at the band edge $(k,E_j^+)$ attains the odd-parity for $k=0$ or $\pi$, 
then $\xi_{R}(E) <0$, and $\xi_R(E) \to 0$ as $E \to E_j^+$ and $\xi_R(E) \to -\infty$ as $E \to E_{j+1}^-$ respectively;
On the other hand,  $\xi_{L}(E) >0$, and $\xi_L \to 0$ as $E \to E_j^+$ and $\xi_L \to +\infty$ as $E \to E_{j+1}^-$ respectively.
\item [(ii)]
If the Bloch edge mode at $(k,E_j^+)$ attains the even-parity, then $\xi_{R}(E) >0$, and $\xi_R(E) \to +\infty$ as $E \to E_j^+$ and $\xi_R(E) \to 0$ as $E \to E_{j+1}^-$
respectively; On the other hand,  $\xi_{L}(E) <0$ and $\xi_L(E) \to -\infty$ as $E \to E_j^+$ and $\xi_L(E) \to 0$ as $E \to E_{j+1}^-$ respectively.
\end{enumerate}
\end{lem}

\noindent\textbf{Proof.}  Without loss of generality, we consider only the case $k=0$ and 
the Bloch mode $\varphi_{j,0}$ at $(0, E_j^+)$ is odd. The proof for other cases is similar. It also suffices to prove for the function $\xi_R(E)$ since the function $\xi_L(E)$ can be treated similarly. 
First, by Lemma \ref{lem-42}, $\varphi_{j,0}(x) =c \psi_{E_j^+, 2}(x)$ for some constant $c$. 
By Lemma \ref{lem-e2}, we have 
$$
\frac{\partial \psi_{E, 2}(1)}{\partial E}(E_j^+) \cdot \psi'_{E_j^+, 2}(1)>0.
$$
On the other hand, note that $\psi_{E_j^+, 2}= \varphi_{j,0} $ is periodic with period one. We have
$$
\psi_{E_j^+, 2}(0)=\psi_{E_j^+, 2}(1)=0, \quad \psi'_{E_j^+, 2}(1) = \psi'_{E_j^+, 2}(0)>0. 
$$ 
Therefore, 
$\frac{\partial \psi_{E, 2}(1)}{\partial E}(E_j^+) >0$, and consequently, we have
$\psi_{E, 2}(1) >0$ for $E \in (E_j^+, E_{j+1}^-)$.

We next define the function 
$$
g(E):=\lambda_{E, 1} - \psi_{E, 1}(1).
$$
By Theorem \ref{thm-parity_change}, $\varphi_{j+1,0}$ is an even state. 
Thus $\varphi_{j+1,0} =c \psi_{E_{j+1}^-, 1}$ for some constant $c$ and we have $\psi_{E_{j+1}^-, 1}(1) = \psi_{E_{j+1}^-, 1}(0) =1$ using the periodicity of $\varphi_{j+1,0}$. It follows that
$$
g(E_{j+1}^-) = \lambda_{E_{j+1}^-, 1} - \psi_{E_{j+1}^-, 1}(1)=1-1=0.
$$
On the other hand, since
$$
\lambda_{E, 1}= \frac{D(E) - \sqrt{D(E)^2 -4}}{2},
$$ 
we have
$$
g'(E) = \frac{1}{2}D'(E) \left(1- \frac{D(E)}{\sqrt{D(E)^2 -4}}\right) - \frac{\partial \psi_{E, 1}(1)}{\partial E}.
$$
It is clear $D(E) \to 2$ and $D'(E)<0$ as $E \to E_{j+1}^-$. Therefore
\be  \label{eq-g'}
\lim_{E \to E_{j+1}^-} g'(E) = \infty,  
\ee  
whence $g(E)<0$ near $E_{j+1}^-$ and hence over the whole interval $(E_j^+, E_{j+1}^-)$. 
This proves that $\xi_R(E) =  \frac{\psi_{E, 2}(1)}{g(E)}<0$  over $(E_j^+, E_{j+1}^-)$. 

We now prove that $\xi_R(E) \to 0$ as $E \to E_j^+$ and $\xi_R(E) \to -\infty$ as $E \to E_{j+1}^-$. 
Recall that $\psi_{E_j^+, 2}(1)=0$. There are two cases: $g(E_j^+)\neq 0$ or $g(E_j^+)=0$. In the former case it is clear that $\xi_R(E) \to 0$ as $E \to E_j^+$. In the latter case, 
$$
\lim_{E \to E_j^+} \xi_R(E)=\lim_{E \to E_j^+} \frac{\psi_{E, 2}(1)}{g(E)} = \lim_{E \to E_j^+} \frac{\frac{\partial \psi_{E, 2}(1)}{\partial E}}{g'(E)} =0, 
$$
where we used the fact that $ \lim_{E \to E_j^+} g'(E) = \infty$ if $g(E_j^+)=0$ (the proof is similar to (\ref{eq-g'})).  
Therefore, in both cases we have $\xi_R(E) \to 0$ as $E \to E_j^+$. 

Finally, we show that $\xi_R(E) \to -\infty$ as $E \to E_{j+1}^-$. Since $g(E_{j+1}^-) = 0$, we need only to show that $\psi_{E_{j+1}^-, 2}(1)\neq 0$.  Indeed, assume otherwise 
$\psi_{E_{j+1}^-, 2}(1)= 0= \psi_{E_{j+1}^-, 2}(0)$. Recall that
$\psi_{E_{j+1}^-, 1}(1) =1$ and that 
$$
\det M(E_{j+1}^-) = \det \begin{pmatrix}
\psi_{E_{j+1}^-, 1}(1) &  \psi_{E_{j+1}^-, 2}(1) \\
\frac{1}{\mu(0)}\psi_{E_{j+1}^-, 1}'(1) &  \frac{1}{\mu(0)} \psi_{E_{j+1}^-, 2}'(1)
\end{pmatrix} = \det 
\begin{pmatrix}
1 &  0 \\
\frac{1}{\mu(0)}\psi_{E_{j+1}^-, 1}'(1) &  \frac{1}{\mu(0)} \psi_{E_{j+1}^-, 2}'(1)
\end{pmatrix}
$$
We have
$$
\frac{1}{\mu(0)} \psi_{E_{j+1}^-, 2}'(1) =1 = \frac{1}{\mu(0)} \psi_{E_{j+1}^-, 2}'(0).
$$
Therefore $\psi_{E_{j+1}^-, 2}$ is a periodic function with period one and hence 
$\varphi_{j+1,0} =c \psi_{E_{j+1}^-, 2}$ 
for some constant $c$. This contradicts to the established fact that $\varphi_{j+1,0} =c \psi_{E_{j+1}^-, 1}$.  This completes the proof of the lemma. \qed

\subsubsection{Interface modes induced by bulk topological indices}
We consider a photonic system which consists of two semi-infinite periodic structures in the left and right half spaces respectively.
Let $\varepsilon_j$ and $\mu_j$ ($j=1, 2$) be the physical parameters for the two periodic structures. 
Both $\varepsilon_j$ and $\mu_j$ are piecewise continuous and real-valued periodic functions with period one and satisfy
$\varepsilon_j(x)= \varepsilon_j(1-x)$, $\mu_j(x) = \mu_j(1-x)$.
The corresponding periodic differential operator is 
$$
 \Lo_j\psi = -\frac{1}{\varepsilon_j(x)} \dfrac{d}{dx} \left( \frac{1}{\mu_j(x)}\frac{d\psi}{dx}\right), \quad j=1, 2.
$$
Then the differential operator for the joint structure is given by
\begin{equation}\label{eq-optA}
\Ao \psi (x) = 
\begin{cases}
 \Lo_1\psi (x) , \quad x <0, \\ 
 \Lo_2\psi (x) , \quad x >0.
 \end{cases}
\end{equation}
\begin{defn}
$\psi$ is called an interface mode of the structure associated with the operator $\Ao$ if $\psi \in L^2(\R)$ satisfy $(\Ao -E)\psi (x) =0$ for some real number $E$. $E$ is called the energy level of the mode $\psi$.
\end{defn}
We investigate the existence of interface modes for the operator $\Ao$. 
In particular, such interface modes are localized near the interface $x=0$ and decay exponentially away from the interface. 
In what follows, we denote the quantities associated with the operator $\Lo_j$ using the superscript $j$ ($j=1, 2$), such as 
the energy level $E_{m}^{(j)}$, the Bloch mode $\psi_{m,k}^{(j)}$, etc.


\begin{thm}
Assume that the following holds:
\begin{enumerate}
\item[(i)]
The operators $\Lo_1$ and $\Lo_2$ attain a common band gap
$$
I:=( E_{m_1}^{(1),+}, E_{m_1+1}^{(1),-}) \cap (E_{m_2}^{(2),+}, E_{m_2+1}^{2,-}) \neq \emptyset
$$
for certain positive integers $m_1$ and $m_2$. 
\item[(ii)]
The bulk topological indices  $\gamma_{m_1}^{(1)} \neq \gamma_{m_2}^{(2)}$ for the operator $\Lo_1$ and $\Lo_2$,
\end{enumerate}
Then there exists an interface mode for the operator $\Ao$. 
In addition, the number of interface modes are given by the number of roots to the equation 
\be
\xi(E):=\xi_{L}^{(1)}(E) - \xi_{R}^{(2)}(E)=0 \quad \mbox{for} \; E\in I.
\ee 
\end{thm}

\noindent\textbf{Proof.}  By Proposition \ref{prop-gap_mode_inv_sym}, $u$ is an interface mode of $\Lo$ at energy level $E$ if and only if $U(0)=c_1 V_{E,2}^{(1)}=c_2 V_{E,1}^{(2)}$
for some constant $c_1, c_2$, or equivalently,
$$
\xi(E):=\xi_{L}^{(1)}(E) - \xi_{R}^{(2)}(E)=0.
$$

Without loss of generality, we consider the case when the common band gap of the operators $\Lo_1$ and $\Lo_2$ is given by $I=(E_{m_1}^{(1),+}, E_{m_1+1}^{(1),-})$.
Moreover,  $\gamma_{m_1}^{(1)}=1$ and $\gamma_{m_2}^{(2)}=-1$ for the two operators.
Then the Bloch mode $\psi_{m_1,k}^{(1)}$ at the band edge $(k,E_{m_1}^{(1),+})$ for the operator $\Lo_1$
is even while the Bloch mode $\psi_{m_2,k}^{(2)}$ at the band edge $(k,E_{m_2}^{(2),+})$ for the operator $\Lo_2$  is odd.
By Lemma \ref{lem-edgemode2},   $\xi_L^{(1)}(E) <0$ and $\xi_L^{(1)}(E) \to -\infty$ as $E \to E_j^+$ and $\xi_L^{(1)}(E) \to 0$ as $E \to E_{j+1}^-$ respectively. 
On the other hand, $\xi_R^{(2)}(E) <0$ and $\xi_R^{(2)}(E) \to 0$ as $E \to E_j^+$ and $\xi_R^{(2)}(E) \to -\infty$ as $E \to E_{j+1}^-$ respectively.
Therefore, for $E$ in the common gap $I$, we see that 
$\xi(E)>0$ for $E$ near $E_{m_1}^{(1),+}$ and $\xi(E)<0$ for $E$ near $E_{m_1+1}^{(1),-}$. 
It follows that there exists a root over the interval $I$ for $\xi(E)=0$.    \qed \\

\begin{rem}
From the proof of the above theorem, one can see that the total number of interface modes (which equal to the number of roots of the function $\xi(\cdot)$) is odd in the generic case. This can be viewed as a form of bulk-interface correspondence since bulk index takes only two values and is in the $\mathbf{Z}_2$ class. 
\end{rem}

\subsection{Stability of interface modes}
Consider a photonic system of the form \eqref{eq-optA}
and $\Ao$ attains an interface mode over a common spectral band gap $I$ of two operators $\Lo_1$ and $\Lo_2$.
Assume that the structure is perturbed locally with a defect region $(d_1, d_2)$, in which $d_1<0<d_2$, and the relative permittivity and permeability of the structure attain the following values:
$$
\varepsilon(x)= 
\begin{cases}
\varepsilon_1(x-d_1), \quad x<d_1,\\
\varepsilon_d(x), \quad d_1<x<d_2, \\
\varepsilon_2(x-d_2), \quad x>d_2.
\end{cases}
\quad \mbox{and} \quad
\mu(x)= 
\begin{cases}
\mu_1(x-d_1), \quad x<d_1,\\
\mu_d(x), \quad d_1<x<d_2, \\
\mu_2(x-d_2), \quad x>d_2.
\end{cases}
$$
We denote the differential operator for the perturbed system by $\Ao_d$, and
denote $$\Psi(x;E)=[\psi(x), \frac{1}{\mu} \psi'(x)]^T,$$
where $\psi$ solves the differential equation $(\Ao_d - E) \psi =0$.

Let 
$$
V_{E,1}^{(j)} =  \begin{pmatrix}
\psi_{E, 2}^{(j)}(1)\\
\lambda_{E, 1}^{(j)} - \psi_{E, 1}^{(j)}(1)
\end{pmatrix}  \quad \mbox{and} \quad
V_{E,2}^{(j)} = \begin{pmatrix}
\psi_{E, 2}^{(j)}(1)\\
\lambda_{E, 2} ^{(j)} - \psi_{E, 1}^{(j)}(1)
\end{pmatrix}
$$ 
be the eigenvectors of the transfer matrix $M^{(j)}(E)$ as defined in \eqref{eq-eigenvec}.
For each $E\in I$, we normalize the eigenvectors $V_{E,1}^{(j)}$ and $V_{E,2}^{(j)}$ by letting $\tilde V_{E,1}^{(j)} = V_{E,1}^{(j)}/\|V_{E,1}^{(j)}\|$ and $\tilde V_{E,2}^{(j)} = V_{E,2}^{(j)}/\|V_{E,2}^{(j)}\|$ and extend them continuously over the closure of the interval $I$. 
Let $M_d(E)$ be the transfer matrix over the defect region $(d_1, d_2)$ such that $\Psi(d_2;E)=M_d(E)\Psi(d_1;E)$.
We see that the localized state is retained for the perturbed system if and only if
\begin{equation}\label{eq-cond_int_perturb}
M_d(E) \tilde V_{E,2}^{(1)}  = c \tilde V_{E,1}^{(2)}
\end{equation}
holds for certain $E\in I$ and some nonzero real number $c$. 
A natural question is how large perturbation is allowed for the defect medium parameters
so that the condition \eqref{eq-cond_int_perturb} holds and the interface mode persists 
for the operator $\Ao_d$. 

\begin{thm}\label{thm-stability1}
Assume that $\Lo_1$ and $\Lo_2$ attain the same band gap $I:=(E_{m_1}^{(1),+}, E_{m_1}^{(1),-}) = (E_{m_2}^{(2),+}, E_{m_2+1}^{2,-})$ 
and the bulk topological indices  $\gamma_{m_1}^{(1)}$ and $\gamma_{m_2}^{(2)}$ are different for the two operators. If
\begin{equation}\label{eq-cond_stablity}
\max{ \Big\{\| \mu \|_{L^\infty(d_1,d_2)}, E \| \varepsilon \|_{L^\infty(d_1,d_2)} \Big\} }  \cdot (d_2-d_1) < \frac{\pi}{2}    
\end{equation} 
holds for any $E \in I$,
then the operator $\Ao_d$ attains an interface mode.
\end{thm}

To prove the theorem, we express the solution vector $\Psi$ as
$$
\Psi(x;E) = \rho \, [\sin\theta, \cos\theta ]^T =
\rho \, [\cos\tilde\theta, \sin\tilde\theta ]^T,
$$
in which the polar angle $\tilde\theta:=\frac{\pi}{2}-\theta$ represents the angle between the $x$-axis and the vector $\Psi$ on the plane.
The radius $\rho$ 
and the angle $\theta$ 
are called Pr\"{u}fer radius and angle, respectively \cite{brown-2013}. Both $\theta$ and $\tilde\theta$ are
unique up to an additive constant integer multiple of $2\pi$.
By a direct calculation, $\rho$, $\theta$ and $\tilde\theta$ satisfy the following equations:
\begin{eqnarray}
(\log\rho)' &=& \frac{1}{2}(\mu-E\varepsilon)\sin(2\theta), \label{eq-rho} \\
\theta' &=& \mu \cos^2\theta + E \varepsilon \sin^2\theta, \label{eq-theta} \\
\tilde\theta' &=& -\mu \sin^2\tilde\theta - E \varepsilon \cos^2\tilde\theta. \label{eq-ttheta}
\end{eqnarray}
In what follows, we view $\rho$, $\theta$ and $\tilde\theta$ as functions of $x$ and $E$.

\begin{lem}\label{lem-Prufer1}
Let $\theta(x_0;E)=\theta_0$, then for any fixed $E>0$, the Pr\"{u}fer angle $\theta(x;\cdot)$ is an increasing function and the polar angle $\tilde\theta(x;\cdot)$ is a decreasing function.
\end{lem}
This is obvious by noting that $\theta'\ge0$ and $\tilde\theta'\le0$. 
Hence the solution vector $\Psi$ rotates clockwisely as $x$ increases for fixed $E$.

\begin{lem}\label{lem-Prufer2}
Let $\theta_1(x;E_1)$ and $\theta_2(x;E_2)$ be the Pr\"{u}fer angle of the solution vector with the energy $E_1 \le E_2$ respectively.
If $\theta_1(x_0;E_1)\le\theta_2(x_0;E_2)$, then $\theta_1(x;E_1)\le\theta_2(x;E_2)$ for all $x>x_0$.
\end{lem}

\begin{lem}\label{lem-Prufer3}
Let $\theta(x_0;E)=\theta_0$, then for any $x>x_0$, $\theta(x;E)$ is an increasing function of $E$.
\end{lem}
The proofs of Lemmas \ref{lem-Prufer2} and \ref{lem-Prufer3} can be found in Corollary 2.3.2 and Theorem 2.3.3 of \cite{brown-2013}. \\

\noindent\textbf{Proof of Theorem \ref{thm-stability1}.} Let $I:=(E_1, E_2)$ be the common band gap of the two operators $\Lo_1$ and $\Lo_2$.
Without loss of generality, we assume that $\gamma_{m_1}^{(1)}=1$ and $\gamma_{m_2}^{(2)}=-1$ so that
the Bloch modes $\psi_{m,k}^{(1)}$ and $\psi_{m,k}^{(2)}$ at the band edge $(k,E_{m}^{+})$ for the operator $\Lo_1$ and $\Lo_2$
are even and odd respectively. In view of Lemma \ref{lem-edgemode2}, as $E$ increases from $E_1$ to $E_2$, 
either  $\tilde V_{E,2}^{(1)}$ or  $-\tilde V_{E,2}^{(1)}$ rotates 
from $U_s:=(-1, 0)^T$ to $U_e:=(0, 1)^T$ in the second quadrant.
On the other hand,  either $\tilde V_{E,1}^{(2)}$ or $-\tilde V_{E,1}^{(2)}$ rotates from $U_e$ to $U_s$ in the second quadrant.

We only consider the case when $\tilde V_{E,2}^{(1)}$ rotates from $U_s$ to $U_e$. 
The other scenarios can be proved in a similar fashion. 
If one sets $\Psi(d_1;E)=\tilde V_{E,2}^{(1)}$, then by Lemmas \ref{lem-Prufer1} and \ref{lem-Prufer2}, the vector $\Psi(d_2;E):=M_d(E) \tilde V_{E,2}^{(1)}$ rotates clockwisely as $E$ increases from $E_1$ to $E_2$.
The corresponding Pr\"{u}fer angle $\theta(d_2;E)$ increases continuously.
If \eqref{eq-cond_stablity} holds, it follows that $\theta' \le \frac{\pi}{2(d_2-d_1)}$ for all $x\in(d_1,d_2)$ and  $E\in I$. We obtain
\begin{equation}\label{eq-dtheta}
    \Delta\theta:=\theta(d_2;E)-\theta(d_1;E)<\frac{\pi}{2} \quad\quad \forall E\in I.
\end{equation}
As such $\Psi(d_2;E_1)$ is located in the second quadrant while $\Psi(d_2;E_2)$ is located in the first quadrant. Therefore, the continuity of the Pr\"{u}fer angle $\theta(d_2;E)$ implies that \eqref{eq-cond_int_perturb} holds for certain $E$ in the bandgap as $\tilde V_{E,1}^{(2)}$ or $-\tilde V_{E,1}^{(2)}$ rotates from $U_e$ to $U_s$ in the second quadrant with increasing $E$.

\qed

If the condition \eqref{eq-cond_stablity} is violated, the stability question is more challenging. Here we provide an answer for a special scenario when the defect only consists of one layer.

\begin{thm}\label{thm-stability2}
Assume that $\Lo_1$ and $\Lo_2$ attain the same band gap 
and the bulk topological indices are different for the two operators.
If $\varepsilon_d(x)\equiv \varepsilon_0$ and $\mu_d(x)\equiv \mu_0$ for certain constants $\varepsilon_0$ and $\mu_0$, 
then the operator $\Ao_d$ attains a localized state for any $\varepsilon_0\ge1$, $\mu_0\ge1$, and $d :=d_2-d_1\ge0$.
\end{thm}

\noindent\textbf{Proof.}   Similar to Theorem \ref{thm-stability1}, we assume that the two operators $\Lo_1$ and $\Lo_2$ attain a common band gap $I:=(E_1, E_2)$, and the toplogical indices for the two operators are $1$ and $-1$.
We denote the trajectory of the end point for the solution vector $\Psi(d_2;E)$ by $\gamma$ as $E$ increases from $E_1$ to $E_2$ in the band gap. Since $\tilde V_{E,1}^{(2)}$ or $-\tilde V_{E,1}^{(2)}$ rotates from $U_e:=(0, 1)^T$ to $U_s:=(-1, 0)^T$ in the second quadrant, while the vector $\Psi(d_2,E)$ rotates clockwisely as $E$ increases, we deduce that \eqref{eq-cond_int_perturb} holds as long as
$\gamma$ crosses the $x$ or $y$ axis on the plane. Next we show that this is true for any $\varepsilon_0\ge1$, $\mu_0\ge1$, and $d$ in the defect layer.

Note that either $\tilde V_{E,2}^{(1)}$ or  $-\tilde V_{E,2}^{(1)}$ rotates from $U_s$ to $U_e$ as $E$ increases from $E_1$ to $E_2$, for brevity we only consider the former.
The transfer matrix $M_d$ is explicitly given by
\begin{equation}\label{eq-Md}
 M_d(E)=\left[
\begin{array}{ll}
\cos(\omega n_d \; d)    &    \dfrac{\mu}{\omega n_d}\sin(\omega n_d \; d) \vspace{0.2cm}\\
-\dfrac{\omega n_d}{\mu}\sin(\omega n_d \; d) & \cos(\omega n_d \; d)
\end{array} \right],
\end{equation}
in which $\omega=\sqrt{E}$ and $n_d = \sqrt{\varepsilon_d\mu_d}$.
Let $\tilde V_{E,2}^{(1)}=[ -v_1(E), v_2(E) ]$ with  $v_1(E)\ge0$ and  $v_2(E)\ge0$, and $\omega_d=\omega n_d \; d$, then
\begin{equation}\label{eq-Psi_d2}
\Psi(d_2;E)  = 
v_1(E) \begin{pmatrix}
-\cos(\omega_d)\\
\dfrac{\omega n_d}{\mu}\sin(\omega_d)
\end{pmatrix}
+
v_2(E) \begin{pmatrix}
\dfrac{\mu}{\omega n_d}\sin(\omega_d) \\
\cos(\omega_d)
\end{pmatrix}.
\end{equation}
In particular,
$$\Psi(d_2;E_1)= \big[-\cos(\omega_{d,1}), \dfrac{\omega_1 n_d}{\mu}\sin(\omega_{d,1})\big]^T,  \quad
\Psi(d_2;E_2)= \big[ \dfrac{\mu}{\omega_2 n_d}\sin(\omega_{d,2}), \cos(\omega_{d,2}) \big]^T.  $$
with $\omega_j =\sqrt{E_j}$ and $\omega_{d,j}=\omega_j n_d \; d$ ($j=1, 2$).

Now assume that $\Psi(d_2;E_1)$ lies in the first quadrant with $\omega_{d,1} \in 2n_1\pi + \big[\frac{\pi}{2}, \pi \big]$ for certain integer $n_1\ge0$.
We only need to consider the case when $\Psi(d_2;E_2)$ also lies in the first quadrant. We observe that $\omega_{d,2} \in 2n_2\pi + \big[0, \frac{\pi}{2} \big]$ for certain integer $n_2>n_1$. Note that $\Psi(d_2;E)$ is located in the lower half plane when $\omega \in 2n_1\pi + \big(\pi, \frac{3\pi}{2} \big)$,
and in the left half plane when $\omega \in 2n_1\pi + \big(\frac{3\pi}{2}, 2\pi \big)$.
Thus the trajectory $\gamma$ crosses both the $x$ and (or) $y$ axis.
One can draw the same conclusion if $\Psi(d_2,E_1)$ lies in other quadrants, and the proof is complete. \qed \\

For a generic defect, the existence of interface modes for the perturbed topological structure is not guaranteed when the condition \eqref{eq-cond_stablity} is violated.
Here we construct counter examples when the defect consists of two layers and the interface mode disappears. The permittivity and permeability values in the defect regions are given by
$$
\varepsilon_d(x)= 
\begin{cases}
\varepsilon_{d,1}, \quad d_1<x<d_*, \\
\varepsilon_{d,2}, \quad d_*<x<d_2
\end{cases}
\quad \mbox{and} \quad
\mu_d(x)= 
\begin{cases}
\mu_{d,1}, \quad d_1<x<d_*, \\
\mu_{d,2}, \quad d_*<x<d_2,
\end{cases}
$$
where the constants $\varepsilon_{d,j}$ and $\mu_{d,j}$ $(j=1,2)$ are to be specified in the following.
Similar to the previous discussions, we assume that the operators $\Lo_1$ and $\Lo_2$ attain the same band gap $I:=(E_1, E_2)$ and the bulk topological indices for the two operators are $1$ and $-1$, respectively.
Furthermore, as $E$ increases in the band gap,  the eigenvector $\tilde V_{E,2}^{(1)}$ rotates from $U_s:=(-1, 0)^T$ to $U_e:=(0, 1)^T$ in the second quadrant.

Let $\Psi_s(x;E)$ and $\Psi_e(x;E)$ be the solution vector for the equation $(\Ao_d - E) \psi =0$
with $\Psi_s(d_1;E)=U_s$ and $\Psi_e(d_1;E)=U_e$, respectively. The corresponding polar angles $\tilde\theta_s(x;E)$ and $\tilde\theta_e(x;E)$ satisfy the equation \eqref{eq-ttheta}. Define $\Delta\tilde\theta:=\tilde\theta_s-\tilde\theta_e$, then $\Delta\tilde\theta$ solves the equation
\begin{equation}\label{eq-dttheta}
(\Delta\tilde\theta)' = (E\varepsilon_d-\mu_d) \sin(\tilde\theta_s+\tilde\theta_e) \sin(\Delta\tilde\theta)  \quad \mbox{in} \; (d_1, d_2).    
\end{equation}

First, we choose $\varepsilon_{d,1}$ and $\mu_{d,1}$ such that $E_1\varepsilon_{d,1}-\mu_{d,1}>0$. Note that $\tilde\theta_s(d_1;E_1)+\tilde\theta_e(d_1;E_1)=\frac{3\pi}{2}$. Since both $\tilde\theta_s$ and $\tilde\theta_e$ are decreasing functions of $x$, one can choose $d_*$ such that
\begin{equation}\label{eq-ttheta_sum}
  \tilde\theta_s(d_*;E_1)+\tilde\theta_e(d_*;E_1)=\pi. 
\end{equation}
Noting that $\Delta\tilde\theta(d_1;E_1)=\frac{\pi}{2}$ and using \eqref{eq-dttheta}, it follows that $(\Delta\tilde\theta)'<0$ in $(d_1,d_*)$ and consequently
\begin{equation}\label{eq-ttheta_diff}
0<\Delta\tilde\theta(d_*,E_1)<\frac{\pi}{2}.
\end{equation}
A combination of \eqref{eq-ttheta_sum} and \eqref{eq-ttheta_diff} yields
\begin{equation}\label{eq-ttheta_se}
\frac{\pi}{2}<\tilde\theta_s(d_*;E_1)<\pi \quad \mbox{and} \quad 0<\tilde\theta_e(d_*;E_1)<\frac{\pi}{2}.    
\end{equation}

Next we choose $\varepsilon_{d,2}$ and $\mu_{d,2}$ such that $E_1\varepsilon_{d,2}-\mu_{d,2}<0$. Furthermore, let $d_2$ be a real number such that
\begin{equation}\label{eq-ttheta_s}
    \tilde\theta_s(d_2;E_1)=\frac{\pi}{2}.
\end{equation}
We deduce from \eqref{eq-ttheta_diff} that $0<\tilde\theta_e(d_2;E_1)<\frac{\pi}{2}$, 
since $\frac{\pi}{2}<\tilde\theta_s+\tilde\theta_e<\pi$
and $(\Delta\tilde\theta)'<0$ in $(d_*,d_2)$.
If $E_2-E_1$ is sufficiently small, one can conclude that
\begin{equation}\label{eq-ttheta_e}
0<\tilde\theta_e(d_2;E_2)<\frac{\pi}{2}.
\end{equation}

From \eqref{eq-ttheta_s} and \eqref{eq-ttheta_e}, it is seen that the solution vector $\Psi(d_2;E):=M_d(E)\tilde V_{E,2}^{(1)}$ rotates in the first quadrant for $E$ in the band gap.
On the other hand, the eigenvector $\tilde V_{E,1}^{(2)}$ or $-\tilde V_{E,1}^{(2)}$ rotates from $U_e$ to $U_s$ in the second quadrant. Therefore, the condition \eqref{eq-cond_int_perturb} for the existence of interface modes does not hold for any $E\in I$.

\section{Resonance of the finite topological structure}\label{sec-resonance}
In this section, we consider the topological structure of finite size that is extended over the interval $(N_1, N_2)$,
where $N_1$ is  a negative integer and $N_2$ is a positive integer.  
The structure is periodic on the left and right of the origin respectively.
More precisely, the permittivity $\varepsilon_N(x)$ and permeability $\mu_N(x)$ of the finite structure takes the following form
$$
\varepsilon_N(x)= 
\begin{cases}
\varepsilon_1(x), \quad N_1<x<0,\\
\varepsilon_2(x), \quad 0<x<N_2, \\
1, \quad \mbox{elsewhere}.
\end{cases}
\quad \mbox{and} \quad
\mu_N(x)= 
\begin{cases}
\mu_1(x), \quad N_1<x<0,\\
\mu_2(x), \quad 0<x<N_2, \\
1, \quad \mbox{elsewhere}.
\end{cases}
$$
in which $\varepsilon_j$ and $\mu_j$ ($j=1, 2$) are piecewise continuous periodic functions with period one.
The corresponding differential operator is 
$$
\Ao_N \psi = -\frac{1}{\varepsilon_N(x)} \dfrac{d}{dx} \left( \frac{1}{\mu_N(x)}\frac{d\psi}{dx}\right).
$$

When an incident wave $\psi^{inc}=e^{ i\omega x}$ impinges from the left of the structure, where $\omega$ is the frequency,
the structure gives rise to the transmitted field $\psi^{tran}= t(\omega) \, e^{i\omega x}$
and the reflected field  $\psi^{ref}=r(\omega) \, e^{-i\omega x}$. The total field $\psi=\psi^{inc}+\psi^{ref}$ for $x<N_1$
and $\psi=\psi^{tran}$ for $x>N_2$, and it satisfies
\begin{equation}\label{eq-scat}
(\Ao_N - \omega^2) \psi = 0, \quad N_1<x<N_2.
\end{equation}
The above scattering problem attains a unique solution for real frequency $\omega$. If the resolvent associated with the scattering problem is extended to the whole complex plane
by analytic continuation, it attains complex-valued poles that are called the resonances of the scattering problem, 
and the associated nontrivial solutions are called quasi-normal modes.
Equivalently, the pole $\omega$ and the corresponding quasi-normal mode $\psi$ solve the following homogenous scattering problem when $\psi^{inc}$=0:
\begin{eqnarray}\label{eq-res_prob}
&&  (\Ao_N - \omega^2) \psi = 0, \quad N_1<x<N_2, \label{eq-res_prob1} \\
&& \frac{1}{\mu_N(N_1)}\frac{d\psi (N_1)}{dx} + i\omega \psi(N_1)=0, \label{eq-res_prob2}\\
&& \frac{1}{\mu_N(N_2)}\frac{d\psi (N_2)}{dx} - i\omega \psi(N_2)=0, \label{eq-res_prob3}
\end{eqnarray}
The last two conditions are outgoing waves conditions imposed on the boundary of the structure.
They are obtained by the continuity of the field across the boundary and
the fact that the outgoing wave takes the form $\psi=c_-e^{-i\omega x}$ and $\psi=c_+e^{i\omega x}$ for $x<N_1$ and $x>N_2$ respectively.

\begin{lem}
Let $\omega\in\mathbb{C}\backslash\{0\}$ be a resonance of \eqref{eq-res_prob1} - \eqref{eq-res_prob3}, then $\omega$ attains negative imaginary part.
\end{lem}

\noindent\textbf{Proof.}  Multiply the differential equation in \eqref{eq-res_prob1} by $\varepsilon\bar\psi$ and integrate by part, it follows
that
\begin{equation*}
\int_{N_1}^{N_2}\dfrac{1}{\mu_N} \abs{\dfrac{d\psi}{dx}}^2 - \omega^2\varepsilon_N\abs{\psi}^2 dx + \dfrac{1}{\mu_N}\dfrac{d\psi(N_1)}{dx}\bar\psi(N_1) 
-\dfrac{1}{\mu_N}\dfrac{d\psi(N_2)}{dx}\bar\psi(N_2) = 0.
\end{equation*}
An application of the boundary conditions yields
\begin{equation}\label{eq-ibp}
\int_{N_1}^{N_2} \dfrac{1}{\mu_N} \abs{\dfrac{d\psi}{dx}}^2 - \omega^2\varepsilon\abs{\psi}^2 dx - i \omega \abs{\psi(N_1)}^2 - i \omega \abs{\psi(N_2)}^2 = 0.
\end{equation}

Let $\omega=\omega_1+i\omega_2$, where  $\omega_1$ and $\omega_2$ are real numbers. 
First let us consider the case when the real part $\omega_1\neq0$. Note that the imaginary part of the left hand side of (\ref{eq-ibp}) is
\begin{equation*}
-\omega_1 \left( 2\omega_2\int_0^a \abs{\psi}^2 dx + \abs{\psi(N_1)}^2 + \abs{\psi(N_2)}^2 \right).
\end{equation*}
If $\omega_2\ge0$, then $\psi(N_1)=\psi(N_2)=0$. This implies that $\dfrac{d\psi(N_1)}{dx}=\dfrac{d\psi(N_2)}{dx}=0$, and consequently 
$\psi\equiv0$ in $(N_1, N_2)$. Hence, we deduce that $\omega_2<0$.
Now if the real part $\omega_1=0$, the left hand side of (\ref{eq-ibp}) is
\begin{equation*}
\int_{N_1}^{N_2} \dfrac{1}{\mu_N} \abs{\dfrac{d\psi}{dx}}^2 + \omega_2^2\abs{\psi}^2 dx +\omega_2\abs{\psi(N_1)}^2+\omega_2\abs{\psi(N_2)}^2=0.
\end{equation*}
If $\omega_2>0$, then a similar argument shows that $\psi\equiv0$ in $(N_1, N_2)$. The proof is complete.  \qed

\medskip
 
We denote the differential operator for the infinite structure (namely when $|N_1|=N_2=\infty$) by
$$
\Ao_\infty \psi (x) = 
\begin{cases}
 \Lo_1\psi (x) , \quad x <0, \\ 
 \Lo_2\psi (x) , \quad x >0,
 \end{cases}
$$
where $ \Lo_j$ ($j=1,2$) is the differential operator with the physical parameters $\varepsilon_j$ and $\mu_j$.
Assume that the structure attains an interface mode $\psi_\infty$ with the energy $E_\infty$.
From the discussions in Sections \ref{sec:perturb_sys} and \ref{sec:inv_sym}, $E_\infty$ is located in a common spectral band gap of two operators.
We would like to investigate resonances for the finite structure that are near the eigenvalue  $\omega_\infty = \sqrt E_\infty$.

Here and henceforth, we set $E=\omega^2$ and let $M^{(j)}(E)$ be the transfer matrix associated with the equation $(\Lo_j-E)\psi=0$.
$\lambda_{E, 1}^{(j)}$ and $\lambda_{E, 2}^{(j)}$ are the eigenvalues of  $M^{(j)}(E)$ defined by \eqref{eq-lambda1_lambda2},
with the corresponding eigenvectors $V_{E,1}^{(j)}$ and $V_{E,2}^{(j)}$ given in \eqref{eq-eigenvec}.
Note that $E_\infty$ is located in the common spectral band gap of $\Lo_1$ and $\Lo_2$,
there holds  $|\lambda_{E, 1}^{(j)}| < 1 < |\lambda_{E, 2}^{(j)}|$ for $E$ in the neighborhood of $E_\infty$.
Without loss of generality, it is assumed that $\psi_{E, 2}^{(j)}(1)\neq0$ so that
the two eigenvectors $V_{E,1}^{(j)}$ and $V_{E,2}^{(j)}$ defined above are linearly independent.
We have the following lemma for the eigenvectors $V_{E,1}^{(j)}$ and $V_{E,2}^{(j)}$.

\begin{lem}\label{lem-dV}
Let $E_0=\omega_0^2$ for $\omega_0\in\mathbb{R}^+$, and $\psi_{E_0, 1}^{(j)}(1)$ and $\psi_{E_0, 2}^{(j)}(1)$ are analytic at $\omega_0$ over the complex plane. 
 If $|\lambda_{E_0,1}^{(j)}| < 1 < |\lambda_{E_0,2}^{(j)}|$, then
$V_{E_0,1}^{(j)}$ and $V_{E_0,2}^{(j)}$ are analytic at $\omega_0$ over the complex plane. Furthermore, there holds
$\det\big[V_{E_0,1}^{(j)}, \frac{d V_{E_0,1}^{(j)}}{d\omega}\big]>0$ and $\det\big[ V_{E_0, 2}^{(j)}, \frac{dV_{E_0,2}^{(j)}}{d\omega}\big]<0$.
\end{lem}
The proof follow the same lines as Theorem 4.4 in \cite{lin-santosa-13}, and we omit here for conciseness.

\begin{thm}\label{thm_res}
Let $N=\min \{ |N_1|, N_2 \}$. There exists an integer $N_0$ such that for any
$N\ge N_0$, there is complex-valued resonance $\omega$ of \eqref{eq-res_prob1} - \eqref{eq-res_prob3}  in the neighborhood of $\omega_\infty$. 
Furthermore, there holds
\begin{equation*}
\abs{\;\omega-\omega_\infty} \le C e^{-\alpha(\omega) N},
\end{equation*}
in which $C$ is a positive constant independent of $N$ and $\alpha(\omega)>0$ is a function defined in the neighborhood of $\omega$.
\end{thm}

\noindent\textbf{Proof}. Let $\Psi(x)= [ \psi(x), \frac{1}{\mu(x)}\psi'(x) ]^T$ the solution vector, where $\psi$ is the solution of \eqref{eq-res_prob1} - \eqref{eq-res_prob3}
with the complex-valued frequency $\omega$.
Note that $|\lambda_{E, 1}^{(j)}| < 1 < |\lambda_{E, 2}^{(j)}|$ for $\omega$ in the neighborhood of $\omega_\infty$, in which $E=\omega^2$,
one can expand $\Psi(N_1)$ as 
$$ \Psi(N_1) = c_1(\omega)  V_{E,1}^{(1)} + c_2(\omega)  V_{E,2}^{(1)}, $$
where the coefficients $c_1(\omega)$ and $c_2(\omega)$ are
\begin{equation}\label{eq-c1c2}
c_1(\omega)=\dfrac{\det[\Psi(N_1), V_{E,2}^{(1)}]}{\det[V_{E,1}^{(1)}, V_{E,2}^{(1)}]}
\quad\mbox{and}\quad c_2(\omega)=\dfrac{\det
[V_{E,1}^{(1)},\Psi(N_1)]}{\det[V_{E,1}^{(1)}, V_{E,2}^{(1)}]}.
\end{equation}
The field at $x=0$ can be expressed as $\Psi(0) = c_1(\omega) \left(\lambda_{E,1}^{(1)}\right)^{|N_1|}  V_{E,1}^{(1)} + c_2(\omega) \left(\lambda_{E,2}^{(1)} \right)^{|N_1|}  V_{E,2}^{(1)}$.
By decomposing $V_{E,1}^{(1)}$ and  $V_{E,2}^{(1)}$ as
$$  V_{E,1}^{(1)} =  c_{11}(\omega)V_{E,1}^{(2)}  + c_{12}(\omega)V_{E,2}^{(2)}, \quad  V_{E,2}^{(1)} =  c_{21}(\omega)V_{E,1}^{(2)}  + c_{22}(\omega)V_{E,2}^{(2)}, $$
where
\begin{eqnarray}\label{eq-cs}
c_{11}(\omega)=\dfrac{\det[ V_{E,1}^{(1)}, V_{E,2}^{(2)}]}{\det[V_{E,1}^{(2)}, V_{E,2}^{(2)}]}, \quad 
c_{12}(\omega)=\dfrac{\det[V_{E,1}^{(2)},V_{E,1}^{(1)}]}{\det[V_{E,1}^{(2)}, V_{E,2}^{(2)}]},  \\
c_{21}(\omega)=\dfrac{\det[ V_{E,2}^{(1)}, V_{E,2}^{(2)}]}{\det[V_{E,1}^{(2)}, V_{E,2}^{(2)}]}, \quad 
c_{22}(\omega)=\dfrac{\det[V_{E,1}^{(2)},V_{E,2}^{(1)}]}{\det[V_{E,1}^{(2)}, V_{E,2}^{(2)}]},
\end{eqnarray}
it follows that
\begin{eqnarray*}
\Psi(0) &=& \left(c_1(\omega) c_{11}(\omega)  \left(\lambda_{E,1}^{(1)}\right)^{|N_1|}  + c_2(\omega) c_{21}(\omega)  \left(\lambda_{E,2}^{(1)} \right)^{|N_1|}  \right) V_{E,1}^{(2)}  \\
            &&  + \left(c_1(\omega) c_{12}(\omega)  \left(\lambda_{E,1}^{(1)}\right)^{|N_1|}  +  c_2(\omega)  c_{22}(\omega)  \left(\lambda_{E,2}^{(1)} \right)^{|N_1|}  \right) V_{E,2}^{(2)}.
\end{eqnarray*}                   
We deduce that the filed at $x=N_2$ is
\begin{eqnarray*}
\Psi(N_2) &=& \left(c_1(\omega)  c_{11}(\omega)   \left(\lambda_{E,1}^{(1)}\right)^{|N_1|}  
                      + c_2(\omega)  c_{21}(\omega)   \left(\lambda_{E,2}^{(1)} \right)^{|N_1|}  \right) \left(\lambda_{E,1}^{(2)}\right)^{|N_2|}  V_{E,1}^{(2)} \\
                 &+&  \left(c_1(\omega)  c_{12}(\omega)   \left(\lambda_{E,1}^{(1)}\right)^{|N_1|}  +  
                 c_2 (\omega)  c_{22}(\omega)  \left(\lambda_{E,2}^{(1)} \right)^{|N_1|}  \right) \left(\lambda_{E,2}^{(2)}\right)^{|N_2|}  V_{E,2}^{(2)} 
                 = \begin{pmatrix}
             \psi(N_2) \\
             i\omega    \psi(N_2)
\end{pmatrix}. 
\end{eqnarray*}
This leads to the equation
\begin{eqnarray}\label{eq-res1}
&& \left(c_1(\omega)  c_{11}(\omega)   \left(\frac{\lambda_{E,1}^{(1)}}{\lambda_{E,2}^{(1)}} \right)^{|N_1|} \left(\frac{\lambda_{E,1}^{(2)}}{\lambda_{E,2}^{(2)}} \right)^{N_2} 
 + c_2(\omega)  c_{21}(\omega)   \left(\frac{\lambda_{E,1}^{(2)}}{\lambda_{E,2}^{(2)}} \right)^{N_2}   \right) \cdot
\left(\lambda_{E, 1}^{(2)} - \psi_{E, 1}^{(2)}(1) - i\omega \psi_{E, 2}^{(2)}(1) \right)  \nonumber \\
&& +  \left(c_1(\omega)  c_{12}(\omega)   \left(\frac{\lambda_{E,1}^{(1)}}{\lambda_{E,2}^{(1)}} \right)^{|N_1|} +  c_2(\omega)   c_{22}(\omega)   \right)
\cdot \left(\lambda_{E, 2}^{(2)} - \psi_{E, 1}^{(2)}(1) - i\omega \psi_{E, 2}^{(2)}(1) \right) = 0,
\end{eqnarray}
which is the equation of resonance.

Let $c(\omega)=\det[V_{E,1}^{(1)}, V_{E,2}^{(1)}] \cdot \det[V_{E,1}^{(2)}, V_{E,2}^{(2)}] $, and define the following complex-valued functions
\begin{eqnarray*}\label{eq-F}
F(\omega) &=&  c(\omega)c_2(\omega)   c_{22}(\omega)  \cdot \left(\lambda_{E, 2}^{(2)} - \psi_{E, 1}^{(2)}(1) - i\omega \psi_{E, 2}^{(2)}(1) \right), \\
G_1(\omega) &=& c(\omega) c_1(\omega)  c_{11}(\omega)  \cdot \left(\lambda_{E, 1}^{(2)} - \psi_{E, 1}^{(2)}(1) - i\omega \psi_{E, 2}^{(2)}(1) \right), \\
G_2(\omega) &=& c(\omega) c_2(\omega)  c_{21}(\omega)   \cdot \left(\lambda_{E, 1}^{(2)} - \psi_{E, 1}^{(2)}(1) - i\omega \psi_{E, 2}^{(2)}(1) \right), \\
G_3(\omega) &=& c(\omega) c_1(\omega)  c_{12}(\omega)  \cdot \left(\lambda_{E, 2}^{(2)} - \psi_{E, 1}^{(2)}(1) - i\omega \psi_{E, 2}^{(2)}(1) \right).
\end{eqnarray*}
The nonlinear equation \eqref{eq-res1} can be written as 
\begin{equation}\label{eq-res2}
G_1(\omega)  \left(\frac{\lambda_{E,1}^{(1)}}{\lambda_{E,2}^{(1)}} \right)^{|N_1|} \left(\frac{\lambda_{E,1}^{(2)}}{\lambda_{E,2}^{(2)}} \right)^{N_2} 
+ G_2(\omega) \left(\frac{\lambda_{E,1}^{(2)}}{\lambda_{E,2}^{(2)}} \right)^{N_2} 
 + G_3(\omega) \left(\frac{\lambda_{E,1}^{(1)}}{\lambda_{E,2}^{(1)}} \right)^{|N_1|}  + F(\omega) = 0.
\end{equation}
Since the infinite structure attains an interface mode $\psi_\infty$ with the energy $E_\infty=\omega_\infty^2$, we have $F(\omega_\infty) = c_{22}(\omega) = 0$.

It can be shown that $F(\omega)$ and $G(\omega)$ are analytic in the neighborhood of the frequency $\omega_\infty$ over the complex plane.
By Taylor's theorem \cite{Ahlfors}, there exists an analytic function $\tilde F(\omega)$ such that
\begin{equation}\label{eq-Taylor}
F(\omega)=\tilde F(\omega)(\omega-\omega_\infty), \quad \mbox{where}  \quad \tilde F(\omega_\infty)=F'(\omega_\infty).
 \end{equation}
Substituting into \eqref{eq-res2} yields
\begin{equation}\label{eq-res3}
 \tilde F(\omega)(\omega-\omega_\infty) = -G_1(\omega)  \left(\frac{\lambda_{E,1}^{(1)}}{\lambda_{E,2}^{(1)}} \right)^{|N_1|} \left(\frac{\lambda_{E,1}^{(2)}}{\lambda_{E,2}^{(2)}} \right)^{N_2} 
- G_2(\omega) \left(\frac{\lambda_{E,1}^{(2)}}{\lambda_{E,2}^{(2)}} \right)^{N_2} 
- G_3(\omega) \left(\frac{\lambda_{E,1}^{(1)}}{\lambda_{E,2}^{(1)}} \right)^{|N_1|}.
\end{equation}
 
Now a direct calculation leads to
$$ F'(\omega_\infty) =  \det[V_{E_\infty,1}^{(1)},\Psi(N_1)]  \cdot \frac{d}{d\omega}\left(\det[V_{E_\infty,1}^{(2)}, V_{E_\infty,2}^{(1)}] \right)  \cdot \left(\lambda_{E_\infty, 2}^{(2)} - \psi_{E_\infty, 1}^{(2)}(1) - i\omega \psi_{E_\infty, 2}^{(2)}(1) \right).$$ 
Noting that $V_{E_\infty,1}^{(2)} = s V_{E_\infty,2}^{(1)}$ for some nonzero constant $s$, we obtain
\begin{eqnarray*}
\frac{d}{d\omega}\left(\det[V_{E_\infty,1}^{(2)}, V_{E_\infty,2}^{(1)}] \right)  &=&  
\det\big[\frac{d V_{E_\infty,1}^{(2)}}{d\omega}, V_{E_\infty,2}^{(1)}\big] + \det\big[V_{E_\infty,1}^{(2)}, \frac{dV_{E_\infty,2}^{(1)}}{d\omega}\big] 
 \\
&=& \frac{1}{s} \det[ \frac{dV_{E_\infty,1}^{(2)}}{d\omega}, V_{E_\infty,1}^{(2)}]  + s \det\big[V_{E_\infty,2}^{(1)}, \frac{dV_{E_\infty,2}^{(1)}}{d\omega}\big].
\end{eqnarray*}
From Lemma \ref{lem-dV} we deduce that there exists a constant $\gamma>0$ such that $\abs{\tilde F(\omega)}\ge\gamma$ in the neighborhood of $\omega_\infty$.
Consequently, we obtain
$$  |\omega-\omega_\infty| \lesssim \max\left\{\left(\frac{\lambda_{E,1}^{(1)}}{\lambda_{E,2}^{(1)}} \right)^{|N_1|}, \left(\frac{\lambda_{E,1}^{(2)}}{\lambda_{E,2}^{(2)}} \right)^{N_2} \right\}
\lesssim  e^{-\alpha(\omega) N},  \quad N=\min \{ |N_1|, N_2 \}, $$
where the last inequality above follows from the fact that $\frac{|\lambda_{E, 1}^{(j)}|}{|\lambda_{E, 2}^{(j)}|}<1$ for $E$ in the neighborhood of $E_\infty$.

If one rewrites the condition \eqref{eq-res3} as $\omega = T(\omega)$, then using 
the inequality $\frac{|\lambda_{E, 1}^{(j)}|}{|\lambda_{E, 2}^{(j)}|}<1$ again, it can be shown that
$T$ is a contraction map in the neighborhood of $\omega_\infty$. Hence the existence of the
resonance follows.   \qed  \\

We illustrate the exponential decay of the distance $\abs{\omega-\omega_\infty}$ by considering a layered period structure.
The structure on the left consists of two layers in each period, with a thickness of $\ell_a^{(1)}=0.42$ and $\ell_b^{(1)}=0.58$ respectively.
The permittivity values of the two layers are $\varepsilon_a^{(1)}=3.8$ and $\varepsilon_b^{(1)}=1$, and the permeability values are $\mu_a^{(1)}=\mu_b^{(1)}=1$.
The structure on the right also consists of two layers in each period, with 
the physical parameters in each period given by $\ell_a^{(2)}=0.38$, $\ell_b^{(2)}=0.62$,
$\varepsilon_a^{(2)}=4.2$, $\varepsilon_b^{(2)}=1$, and $\mu_a^{(2)}=\mu_b^{(2)}=1$. 
When $|N_1|=N_2=\infty$, the structure attains an interface mode at the frequency $\omega_\infty=15.6765$.
Table \ref{tab-res} shows the value $\omega-\omega_\infty$ and $\abs{\omega-\omega_\infty}$ when $|N_1|=N_2= 2, 4, 8, 16$.
It is observed that the distance $\abs{\omega-\omega_\infty}$ decays exponentially with respect to the number of period $N$.

\begin{figure}[!htbp]
\begin{center}
\includegraphics[height=5cm]{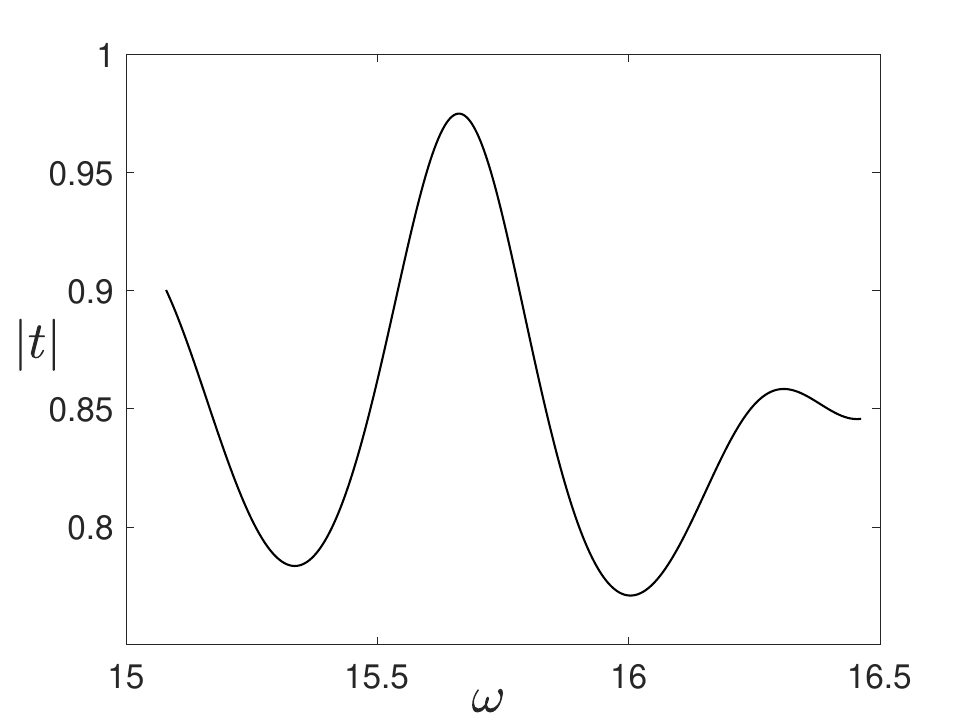}
\includegraphics[height=5cm]{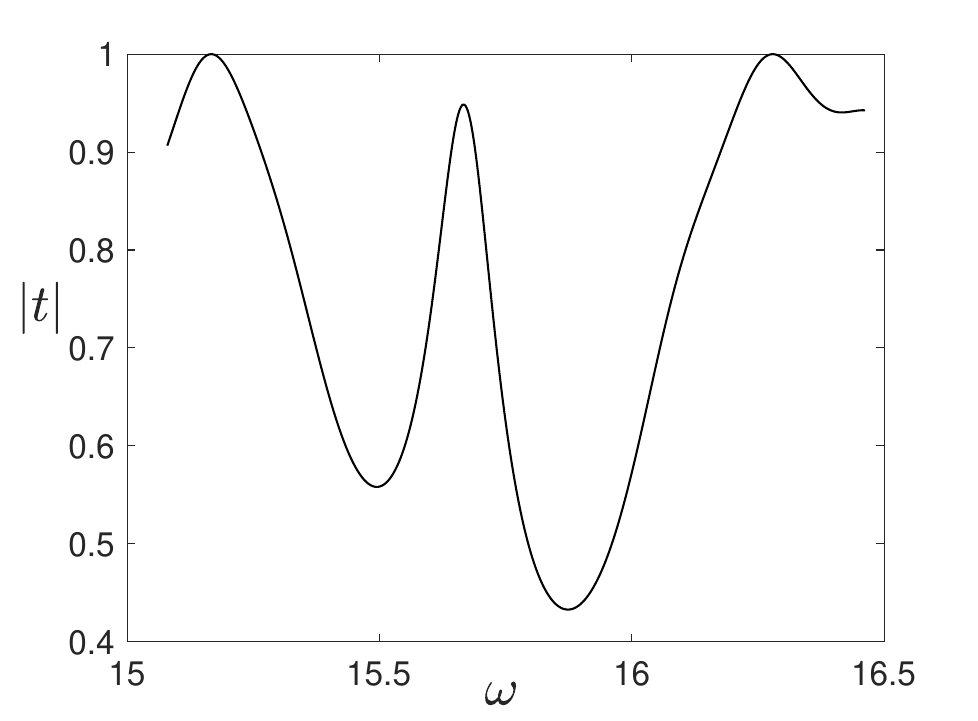}  \\
\includegraphics[height=5cm]{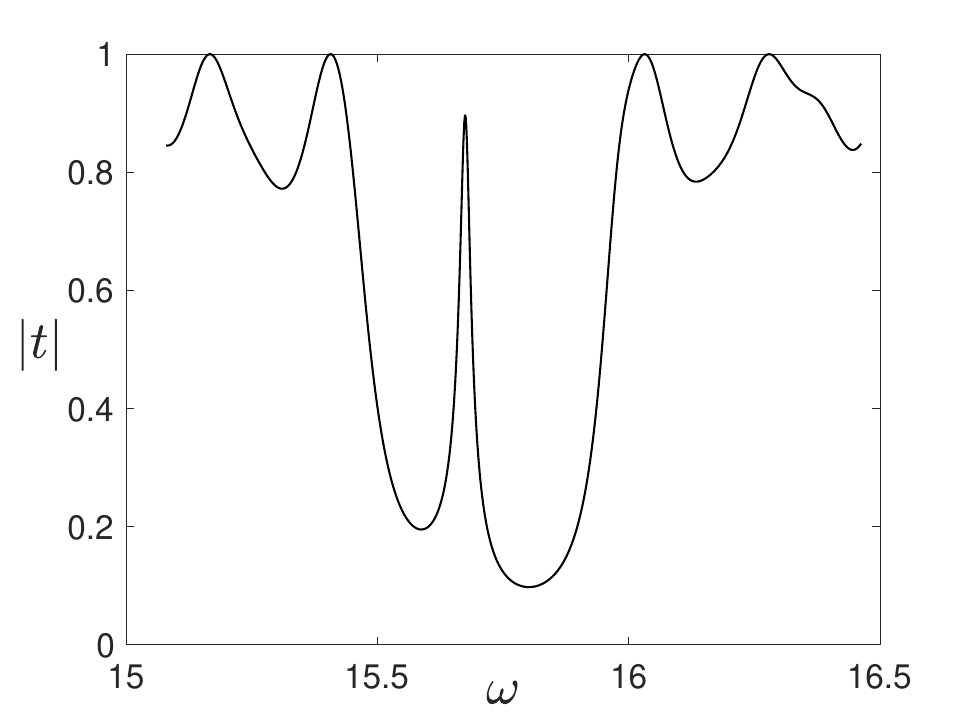}  
\includegraphics[height=5cm]{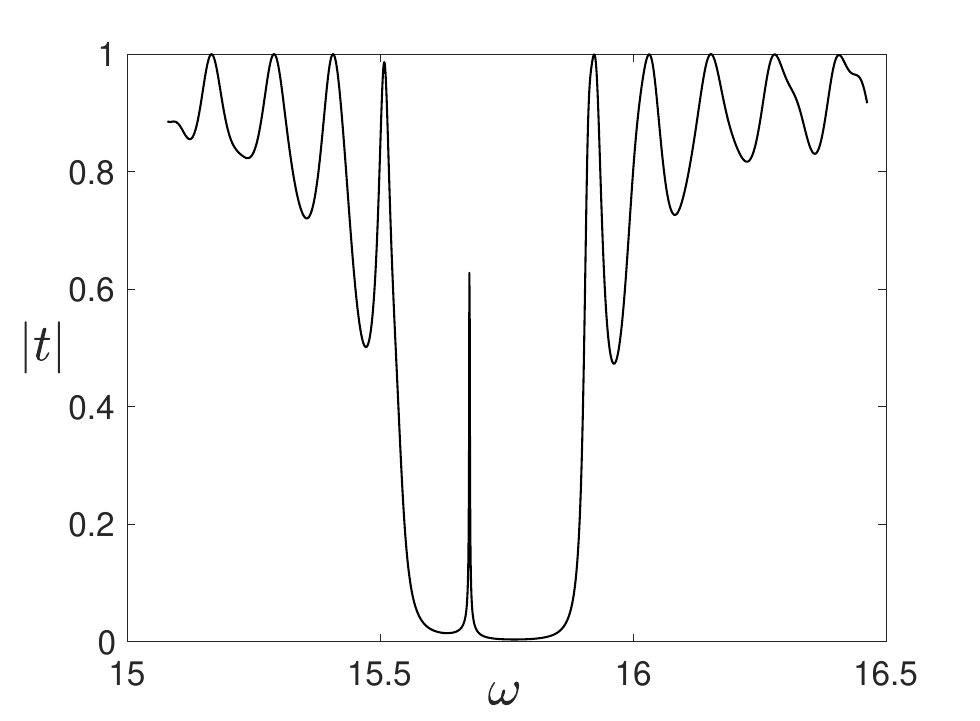}  
\caption{Transmission value $\abs{t}$ near the resonant frequency for $|N_1|=N_2=2, 4, 8, 16$. 
The infinite structure attains an interface mode at the frequency $\omega_\infty=15.6765$. }\label{fig:prob_t_diff_N}
\end{center}
\end{figure}

Now considering the scattering problem \eqref{eq-scat} with the incident wave $\psi^{inc}=e^{ i\omega x}$. The transmission
$|t|$ exhibits peaks at resonant frequencies. As shown in Figure \ref{fig:prob_t_diff_N},
when $N$ increases, the resonant peaks become sharper as the imaginary part of the resonance decreases.

\begin{table}[!htbp]
\begin{center}
\begin{tabular}{ccccc}
  \hline
  $N$ & 2 & 4 & 8 & 16          \\
  \hline\ 
$\Re (\omega-\omega_\infty)$ &  -0.0132 & $ -0.0065 $   & $-0.0016$ &  $-5.81\times 10^{-5}$  \\
$\Im (\omega-\omega_\infty)$ & - 0.2241 & $ - 0.0671$   & $ -0.0104$ &  $- 4.08\times 10^{-4}$  \\
$\abs{\omega-\omega_\infty}$ & 0.2245 & $ 0.0674$   & $ 0.0105$ &  $ 4.12\times 10^{-4}$  \\
\hline
\end{tabular}
\caption{Resonances for finite structures with different number of period.} 
\label{tab-res}
\end{center}
\end{table}

\bibliography{references}

\end{document}